\documentclass[%
 superscriptaddress,
 twocolumn,
 showpacs
nofootinbib,
 amsmath,amssymb,
 prd,
balancelastpage,
nofootinbib,
floatfix
]{revtex4-1}

\usepackage{graphicx}
\usepackage{dcolumn}
\usepackage{bm}
\usepackage{amsmath}
\usepackage{diagbox}
\usepackage{rotating}
\usepackage{multirow}
\usepackage{xcolor}
\usepackage{hepunits}

\usepackage{hyperref}
\usepackage[mathlines]{lineno}
\newcommand{\be}{ \begin{eqnarray} }
\newcommand{\ee}{ \end{eqnarray} }
\newcommand{\vecq}{\mathbf{q}}
\newcommand{\vecv}{\mathbf{v}}
\newcommand{\veck}{\mathbf{k}}
\newcommand{\vecx}{\mathbf{x}}

\newcommand{\db}[1]{ \textcolor{red}{(DB: #1)} }

\begin{document}

{\small \hfill FERMILAB-PUB-20-065-A}

\title{A Dark Matter Interpretation of Excesses in Multiple Direct Detection Experiments}

\author{Noah~Kurinsky}\thanks{kurinsky@fnal.gov}
\affiliation{Fermi National Accelerator Laboratory, Batavia, Illinois 60510, USA}
\affiliation{Kavli Institute for Cosmological Physics, University of Chicago, Chicago, Illinois 60637, USA}

\author{Daniel~Baxter}\thanks{dbaxter@kicp.uchicago.edu}
\affiliation{Kavli Institute for Cosmological Physics, University of Chicago, Chicago, Illinois 60637, USA}
\affiliation{Enrico Fermi Institute, University of Chicago, Chicago, Illinois 60637, USA}

\author{Yonatan~Kahn}\thanks{yfkahn@illinois.edu}
\affiliation{University of Illinois at Urbana-Champaign, Urbana, Illinois 61801, USA}

\author{Gordan~Krnjaic}\thanks{krnjaicg@fnal.gov}
\affiliation{Fermi National Accelerator Laboratory, Batavia, Illinois 60510, USA}
\affiliation{Kavli Institute for Cosmological Physics, University of Chicago, Chicago, Illinois 60637, USA}

\date{\today}

\begin{abstract}
We present a novel unifying interpretation of excess event rates observed in several dark matter direct-detection experiments that utilize single-electron threshold semiconductor detectors. Despite their different locations, exposures, readout techniques, detector composition, and operating depths, these experiments all observe statistically significant excess event rates of $\sim$ 10 Hz/kg. 
However, none of these persistent excesses has yet been reported as a dark matter signal because individually, each can be attributed to different well-motivated but unmodeled backgrounds, and taken together, they cannot be explained by dark matter particles scattering elastically off detector nuclei or electrons.
We show that these results can be reconciled if the semiconductor detectors are seeing a collective inelastic process, consistent with exciting a plasmon. We further show that plasmon excitation could arise in two compelling dark matter scenarios, both of which can explain rates of existing signal excesses in germanium and, at least at the order of magnitude level, across several single-electron threshold detectors. At least one of these scenarios also yields the correct relic density from thermal freeze-out. Both dark matter scenarios motivate a radical rethinking of the standard interpretations of dark matter-electron scattering from recent experiments.

\end{abstract}

\maketitle


\section{Introduction}

\begin{figure}[t]
    \centering
    \includegraphics[width=0.5\textwidth,trim=20 0 80 0,clip=true]{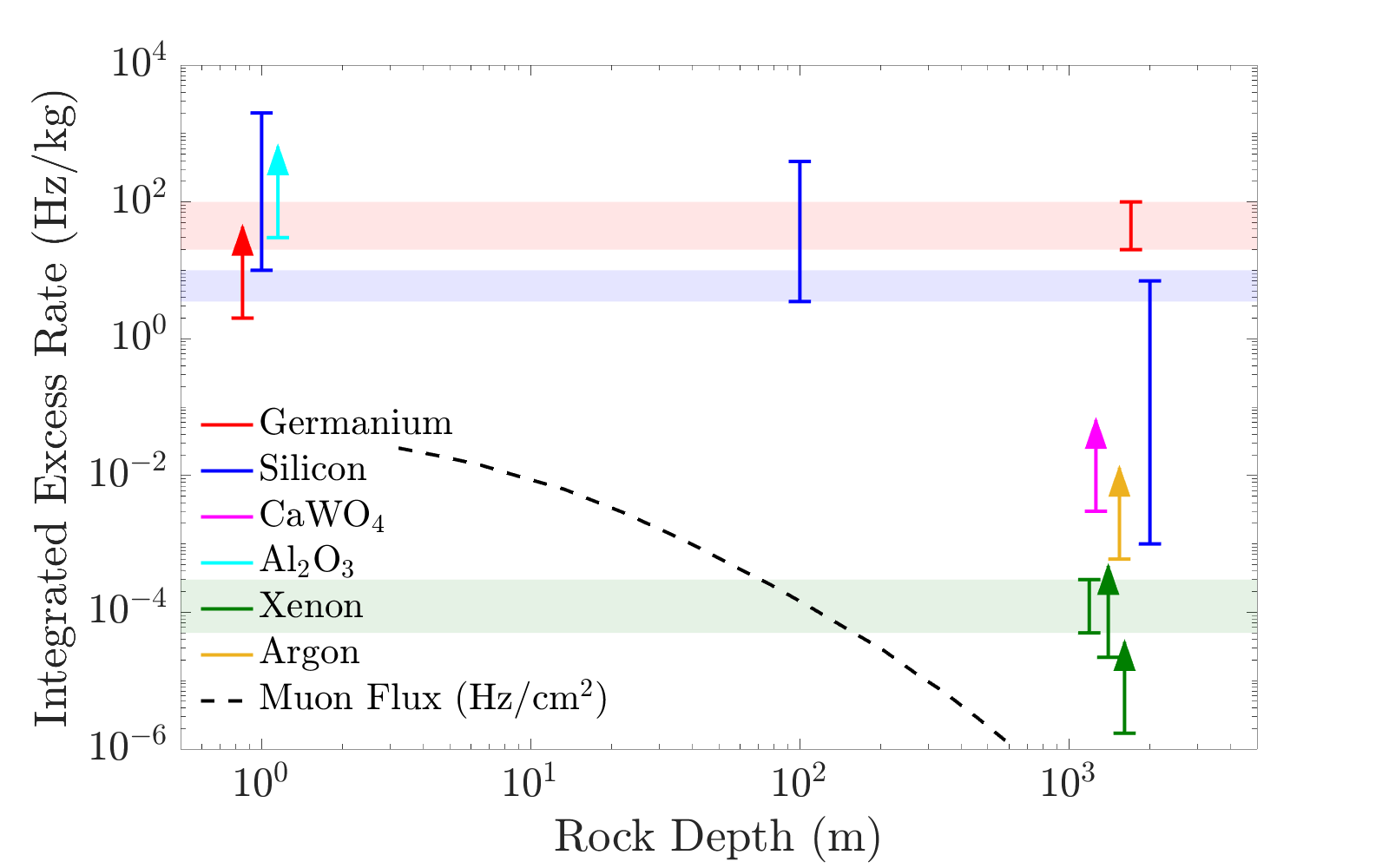}
    \caption{ Integrated rate of each excess versus approximate depth (shifted for clarity), separated by detector medium. Ranges are given according to the same criteria in Table~\ref{tab:eventrates} with the shaded bands indicating regions most consistent with all observed excess rates for Ge (red), Si (blue), and Xe (green), along with the muon flux from~\cite{muon_bugaev} in dashed black to highlight the lack of dependence on depth. For the measurements which only give a lower bound on excess rate, we indicate the possibility of a larger total excess rate with an arrow. 
    Given some reasonable model for the spectrum of each excess below threshold, an upper bound would apply to these measurements, but determining such a bound is outside of the scope of this paper. We note that there exists some tension among the silicon measurements shown here, indicative of non-negligible unresolved detector backgrounds which are not in conflict with the order of magnitude arguments herein.
    }
    \label{fig:ratevdepth}
\end{figure}

Searches for particle dark matter (DM) with masses below 1~GeV have proliferated in the last decade, driven by advances in detector technologies which have pushed heat detection thresholds below 100~eV \cite{EdelweissWIMP,Abdelhameed_2019} and charge detection thresholds to the single electron-hole pair level \cite{Agnese_2018,Abramoff_2019,Essig_2017}. 
While high-mass ($\gtrsim 1$ GeV) searches have continued to advance to larger background-free exposures, several low-mass searches, including 
EDELWEISS~\cite{EdelweissWIMP,edelweissHV},
CDMS HVeV~\cite{Agnese_2018},
SENSEI~\cite{Abramoff_2019},
DAMIC~\cite{PhysRevLett.123.181802},
CRESST-III~\cite{Abdelhameed_2019},
$\nu$CLEUS~\cite{Angloher_2017},
XENON10~\cite{Essig_2012,Essig_2017},
XENON100~\cite{Essig_2017}, 
XENON1T~\cite{xenon1t}, and
Darkside50~\cite{Agnes:2018oej} -- see Table \ref{tab:eventrates} and Fig.~\ref{fig:ratevdepth} -- have observed events at low energy superficially consistent with either dark rate or unmodeled backgrounds. As more experiments approach these low-mass regions, it is pertinent to ask whether these excess rates -- defined as the residual efficiency-corrected rate after subtracting known, modeled backgrounds -- all have independent origins (as is typically assumed), or if a single mechanism can provide a unifying explanation.


The standard signal interpretation of an excess in a detector with order 100~eV threshold is that of elastic nuclear recoils from DM (as described by \citet{LEWIN199687}), whereas a detector with a single-electron threshold is considered primarily sensitive to DM scattering on electrons (as described in detail by~\citet{Essig:2015cda} for semiconductors, see also \cite{Essig:2011nj,Graham:2012su,Lee:2015qva} for earlier work). 
As has been recently shown in Refs.~\cite{Baxter:2019pnz,Essig:2019xkx}, the lines between these interpretations blur in the case of inelastic below-threshold nuclear recoils with accompanying above-threshold ionization, which, for a liquid noble detector, can be the dominant signal component for DM masses between approximately 100--1000~MeV. Because the term ``inelastic'' has different meanings in the theoretical and experimental communities, we emphasize that in this paper, ``inelastic'' refers to an energy and momentum transfer to the \emph{detector} which differs from the relations from two-to-two scattering. In particular, it refers to exciting internal modes of the detector, \emph{not} internal modes of the DM.



In this paper, we postulate that existing excesses in silicon (Si), germanium (Ge), and sapphire (Al$_2$O$_3$) detectors can be persuasively interpreted as the excitation of a plasmon resonance, a ubiquitous feature of nearly every well-ordered solid-state material \cite{raether2006excitation}. The strong plasmon resonance in highly ordered crystals, and the absence of such a resonance in less ordered materials, provides a natural explanation for the large rate differences observed between these detectors and other materials such as CaWO$_4$ and liquid xenon or argon. Indeed, plasmon excitation is the quintessential many-body effect, and provides an important example of an inelastic process that dominates at \emph{low} momentum transfers and which \emph{cannot} be understood in terms of two-body scattering and non-interacting single-particle states, as has been the standard treatment of DM-electron interactions \cite{Essig:2015cda}. 

We argue that a compelling  explanation of the common~$\sim 10$ Hz/kg event rate seen in numerous charge detection experiments, in widely varying background environments, is lacking if the excesses are attributed to plasmon excitation sourced by known Standard Model (SM) particles.
In contrast, these rates can be explained by a common DM origin, albeit through interactions that primarily excite collective charge modes in well-ordered crystals. We will argue that these interactions are easily accommodated by the most widely-studied DM benchmark models and have simply been neglected in previous studies in favor of the more familiar electron and nuclear recoils. Furthermore, one of these benchmark scenarios can explain the DM cosmological abundance with the same interaction strength that accommodates these experimental excesses.

This paper is organized as follows. In Sec.~\ref{sec:Review}, we review the various excesses in low-threshold experiments and propose a yield model which reconciles the observed ionization ($E_e$)~\cite{edelweissHV} and calorimetric ($E_{det}$)~\cite{EdelweissWIMP} spectra in EDELWEISS germanium data, the only material for which excesses are currently observed in both $E_e$ and $E_{det}$ data. 
In Sec.~\ref{sec:Plasmon}, we show that interpreting the yield model as a plasmon is  consistent with the similar event rates measured in silicon and Al$_2$O$_3$ detectors as well as with the comparably lower rates measured in amorphous materials like CaWO$_4$ and liquid Xe. Additionally, we argue that SM sources cannot produce plasmon excitation rates consistent with the observed excesses. In Sec.~\ref{sec:Models} we present two illustrative DM scenarios which can explain the observed rate in Ge, and demonstrate their consistency with the other excesses at the order-of-magnitude level. Moreover, we point out that the excess rates considered are just beginning to scrape the models explored, while still likely containing some background; a follow-up demonstration of a lower rate by one of the experiments considered here would further probe interesting parameter space for the models we present. 
We conclude in Sec.~\ref{sec:Conclusions} with a number of predictions and suggestions for future studies. Further details on dark counts in semiconductor detectors, yield curves, and plasmons are provided in the Appendices.


\begin{table*}[!th]
    \centering
    \begin{tabular}{|c|c|c|c|c|c|c|c|}
    \hline
    Readout Type& Target & Resolution & Exposure & Threshold & Excess Rate (Hz/kg) & Depth & Reference \\
    \hline
      \multirow{4}{*}{Charge ($E_e$)} & Ge & 1.6~$e^{-}$ & 80 g$\cdot$d & 0.5~eVee ($\sim$1$e^{-}$)\footnote{There is a very small but non-zero sensitivity to single electrons that, when the large exposure is taken into account, becomes comparable in sensitivity to the other electron recoil experiments.}& [20, 100] & 1.7~km & EDELWEISS \cite{edelweissHV}\\ 
      & Si & $\sim$0.2 $e^{-}$ & 0.18 g$\cdot$d & 1.2~eVee ($<$1~$e^{-}$) & [6, 400] & 100~m & SENSEI \cite{Abramoff_2019} \\ 
      & Si & 0.1 $e^{-}$ & 0.5 g$\cdot$d & 1.2~eVee ($<$1~$e^{-}$) & [10, 2000] & $\sim$1~m & CDMS HVeV \cite{Agnese_2018} \\ 
      & Si & 1.6~$e^{-}$ & 200 g$\cdot$d & 1.2~eVee ($\sim$1$e^{-}$) & [1 $\times 10^{-3}$, 7] & 2~km & DAMIC \cite{PhysRevLett.123.181802} \\ 
      \hline
      \multirow{3}{*}{Energy ($E_{det}$)} & Ge & 18~eV & 200 g$\cdot$d & 60~eV & $>$~2 & $\sim$1~m & EDELWEISS \cite{EdelweissWIMP} \\
      & CaWO$_4$ & 4.6~eV & 3600 g$\cdot$d & 30~eV & $>$ 3 $\times 10^{-3}$ & 1.4~km & CRESST-III~\cite{Abdelhameed_2019} \\
      & Al$_2$O$_3$ & 3.8~eV & 0.046 g$\cdot$d & 20~eV & $>30$ & $\sim$1~m & $\nu$CLEUS \cite{Angloher_2017} \\
      \hline
      \multirow{4}{*}{Photo $e^{-}$} 
      & Xe & $6.7$ PE ($\sim 0.25 \, e^{-}$) & 15 kg$\cdot$d & 12.1~eVee ($\sim$14~PE) & [0.5, 3] $\times 10^{-4}$ & 1.4~km & XENON10 \cite{Essig_2012,Essig_2017} \\
      & Xe & $6.2$ PE ($\sim 0.31 \, e^{-}$) & 30~kg$\cdot$yr & $\sim$70~eVee ($\sim$80~PE) & $>2.2$ $\times 10^{-5}$ & 1.4~km & XENON100 \cite{Essig_2017} \\
      & Xe & $<10$ PE & 60~kg$\cdot$yr & $\sim$140~eVee ($\sim$90~PE) & $>$~1.7 $\times 10^{-6}$ & 1.4~km & XENON1T \cite{xenon1t} \\
      & Ar & $\sim$15 PE ($\sim 0.5 \, e^{-}$) & 6780 kg$\cdot$d & 50 eVee
  & $> 6$ $\times 10^{-4}$  & 1.4~km & Darkside50 \cite{Agnes:2018oej} \\
    \hline
    \end{tabular}
    \caption{Rates of observed low-energy excesses in experiments with single electron ($<$100 eV) charge (energy) resolutions. Lower bounds on the rate are given by integrating the rate above 2e$^-$ (or above threshold) whereas upper bounds are given by assuming that the entire 1e$^-$ rate is of the same origin, despite likely containing large experiment-specific backgrounds (see Appendix \ref{app:dc} for a discussion). Experiments sensitive to charge energy $E_e$ are in the top section of the table, while experiments sensitive to total detector energy $E_{det}$ are in the middle section. The bottom section lists experiments sensitive to secondary radiation produced by charge interactions. The main coincidence reported here is that the excesses for $n_e\ge 2$ across the first three charge detectors ($\sim$10 Hz/kg) demonstrate  {\it nearly identical} rates for their $n_e \geq 2$ bins (20, 6, and 10 Hz/kg), despite spanning $\sim$ 2~km of variation in overburden and almost three orders of magnitude in exposure. The total rate observed in the DAMIC detector is much lower, but the upper bound (7 Hz/kg) is intriguingly of the same order of magnitude.}
    \label{tab:eventrates}
\end{table*}

\section{Review of Recent Low-Threshold Results}
\label{sec:Review}

We begin by considering the standard interpretation of existing excesses in roughly chronological order of appearance to illustrate the difficulty in explaining them through conventional backgrounds. We restrict our discussion to experiments running detectors with source-independent energy resolution below 100~eV, where excesses are observed directly.

\subsection{Nuclear Recoil Searches}

In a typical nuclear recoil (NR) search, an excess manifests as an unexplained event rate rising with decreasing energy down to the detector threshold; by contrast most background processes are approximately flat in $E_{det}$ at these energies. Calorimetric detectors are sensitive to $E_{det}$ in the form of phonons, which are the longest-lived excitations after the relaxation of all charge processes.



\medskip

{\bf }
{\bf \noindent CRESST-III:} In the calorimetric energy channel, the first hint of an unexpected signal at low energy came from the CRESST-III experiment~\cite{Abdelhameed_2019}. With a heat threshold of 30~eV in a CaWO$_4$ detector, this was the first result to achieve significant exposure (3.6 kg$\cdot$days) below 100~eV, and their initial hypothesis for the excess of 440 events near threshold was crystal cracking \cite{crystalCracking}. The relatively low rate ($3 \times 10^{-3}$ Hz/kg; see Table~\ref{tab:eventrates}) and the lack of other measured excesses at the time suggested this hypothesis as the least controversial explanation.


\medskip

{\bf  \noindent $\nu$CLEUS:} Shortly thereafter, the $\nu$CLEUS experiment~\cite{Angloher_2017}, an off-shoot of the CRESST collaboration targeting coherent neutrino scattering, published a surface NR search in which an excess of 30 Hz/kg above expected background was observed in an Al$_2$O$_3$ detector. Taken alone, this rate could potentially also be interpreted as crystal cracking \cite{crystalCracking}, though this would require an explanation for the drastically higher rate than the excess observed in CaWO$_4$.


\medskip
{\noindent \bf EDELWEISS:} Most recently, EDELWEISS published surface results from a Ge detector with a 60~eV threshold~\cite{EdelweissWIMP} showing a very large low-energy excess above the expected background. This excess steeply rises below 500~eV and reaches over 100 times the measured flat background rate at the detector threshold of 60~eV, leaving no argument about its statistical significance; they observe on the order of $10^5$ events. The observed rate above threshold is orders of magnitude larger than the CRESST-III excess and extends to higher energy, making it inconsistent with a simultaneous elastic nuclear recoil interpretation of the two. Independently of the CRESST-III excess, the EDELWEISS excess has eluded interpretation as a DM signal, since the sharp rise matches neither the expected spectrum of an elastic DM recoil (using the standard velocity distribution \cite{LEWIN199687}), nor secondary ionization induced by DM-nuclear scattering, the so-called Migdal effect (using \citet{Ibe:2017yqa} to calculate the cross section for this process).

\medskip
The evidence from the $E_{det}$ spectra is thus inconclusive at this point in the story: multiple experiments observe excesses, none consistent with each other, without a unifying explanation (apart from crystal cracking, which should not produce any charge).\footnote{It should be noted that one other experiment running a calorimetric detector has also noted a low energy excess. SuperCDMS, running an 11~g Si detector, has measured an excess above a threshold of around 20~eV \cite{pd2LTD}, but the rate was not published at that time. The rate rises above background around 30~eV, and also appears to be sharply rising.  When information becomes available about the spectrum of this excess, it can be incorporated into this analysis, but at this time it remains a qualitatively interesting result which we cannot interpret further.}

\subsection{Electron Recoil Searches}
In a typical electron recoil (ER) search, dark counts are expected to contribute a significant quantity of single-electron events, and for a given detector should produce a calculable number of pile-up events with two electrons (see Appendix~\ref{app:dc}). An excess rate in ionization energy $E_e$ can thus either be interpreted as the number of events with two electrons exceeding this prediction, or the overall dark rate, interpreted as a limit on a putative signal rate.

\medskip
{\bf\noindent CDMS HVeV/SENSEI: }
The successful demonstration of single-electron thresholds in Si detectors by CDMS HVeV \cite{Agnese_2018} and SENSEI \cite{Abramoff_2019} led to a leap forward in electron recoil sensitivity to low-mass DM. Both experiments observed a roughly $\sim$Hz/g dark rate in the single electron bin, and only ran for less than a gram-day of exposure. The relative similarity of the event rates was striking, but was considered to be a temporary coincidence that would soon be resolved as one of the experiments improved on their single electron dark rates. It is notable that neither experiment has demonstrated an improved dark rate as of this writing, which may point to a dark rate which is independent of detector environment and is not reduced with additional overburden.

\medskip
{\bf\noindent EDELWEISS: }
Subsequently, the first electron recoil analysis in Ge was released by EDELWEISS \cite{edelweissHV}; intriguingly, the observed event rate is within an order of magnitude of the Si rates, despite exposures differing by a factor of 400 among the three experiments, and the fact that the EDELWEISS search was conducted with significantly greater overburden. Further investigation reveals that the event rate per unit mass in the 2--3 electron bins is remarkably similar between the three experiments, the Ge rate being only roughly twice the Si rate.

\medskip
{\bf\noindent DAMIC: }
Finally, the latest DAMIC \cite{PhysRevLett.123.181802} limit is stronger than the other ER limits, as explained by the significantly reduced dark rate in the single electron bin compared to other silicon detectors. 
The ER analysis presented by DAMIC does not have single-electron resolution and instead assumes Poisson-distributed dark counts, from which we extract a robust upper bound on the 1$e^-$ bin and an inferred lower bound on the 2$e^-$ bin.
The DAMIC data is most in tension with the narrative presented here, indicating a source of events in CDMS HVeV and SENSEI that is absent in the DAMIC detector. 
Regardless, we would like to emphasize that the origin of the dark current in DAMIC remains unknown and could still be consistent with some realizations of the interpretation presented here.

\medskip
{\bf\noindent XENON10/100/1T: } At face value, a DM-electron scattering interpretation in the semiconductor detectors is strongly inconsistent with results from XENON10 \cite{Essig_2012}, which sees a far smaller event rate. 
We list observed event rates at the bottom of Table~\ref{tab:eventrates} for several noble liquid experiments with phototube readout; of these, only XENON10 reports a single charge rate because its threshold is below the average energy (13.7~eV) needed to produce one quantum of charge in xenon \cite{dahl:2009phd}. Regardless of any assumption about whether the excesses in the XENON experiments arise from the same source as those in the semiconductor experiments (as we will explore further in Sec.~\ref{sec:SignalOrigin} below), any consistent explanation of the semiconductor excesses must explain the orders of magnitude lower event rates observed in liquid noble experiments.
All of these experiments do still observe unexplained excesses at low energy, as shown in Tab.~\ref{tab:eventrates}.\footnote{We note that a recent result using phototube readout of EJ-301 scintillator reports a total single photoelectron rate of 3.8 Hz, corresponding to a mass-normalized single scintillation photon production rate of 14 Hz/kg \cite{collar_simp,Blanco:2019lrf}, much larger than the noble liquid rates and comparable to the semiconductor rates. However, since this experiment was the first demonstration of a new technique for light DM searches and was run with minimal overburden, we regard this result as qualitatively interesting and await further data from an underground run.}  
A significant amount of work has been put into better understanding the source of these excess event rates in xenon TPCs~\cite{xenonS2_2007,xenonS2_2013,xenonS2_2017,xenonS2_2018}; however, we note that at very least \emph{some} event rate appears to scale with detector mass \cite{Bernstein:2020cpc}, as would be expected from a dark matter signal.




\subsection{Determining Signal Origin}
\label{sec:SignalOrigin}

The significance of the apparent coincidences in the semiconductor detectors is that these detectors acquired data in very different environments (both near surface and deep underground), each with distinct technologies, at dramatically different temperatures and electric fields, with greatly varying degrees of shielding. There is no detector effect or known background that should conspire to produce the same event rate in these detectors. Furthermore, in all four charge-readout detectors, a charge produced with arbitrarily low energy above the band edge may be detected: there is no threshold for charge detection. By contrast, the calorimetric searches have a nonzero energy threshold, below which events can be hidden depending on the energy spectrum of the signal. 

\begin{figure*}
    \centering
    \begin{minipage}{0.45\textwidth}
    \includegraphics[width=\textwidth]{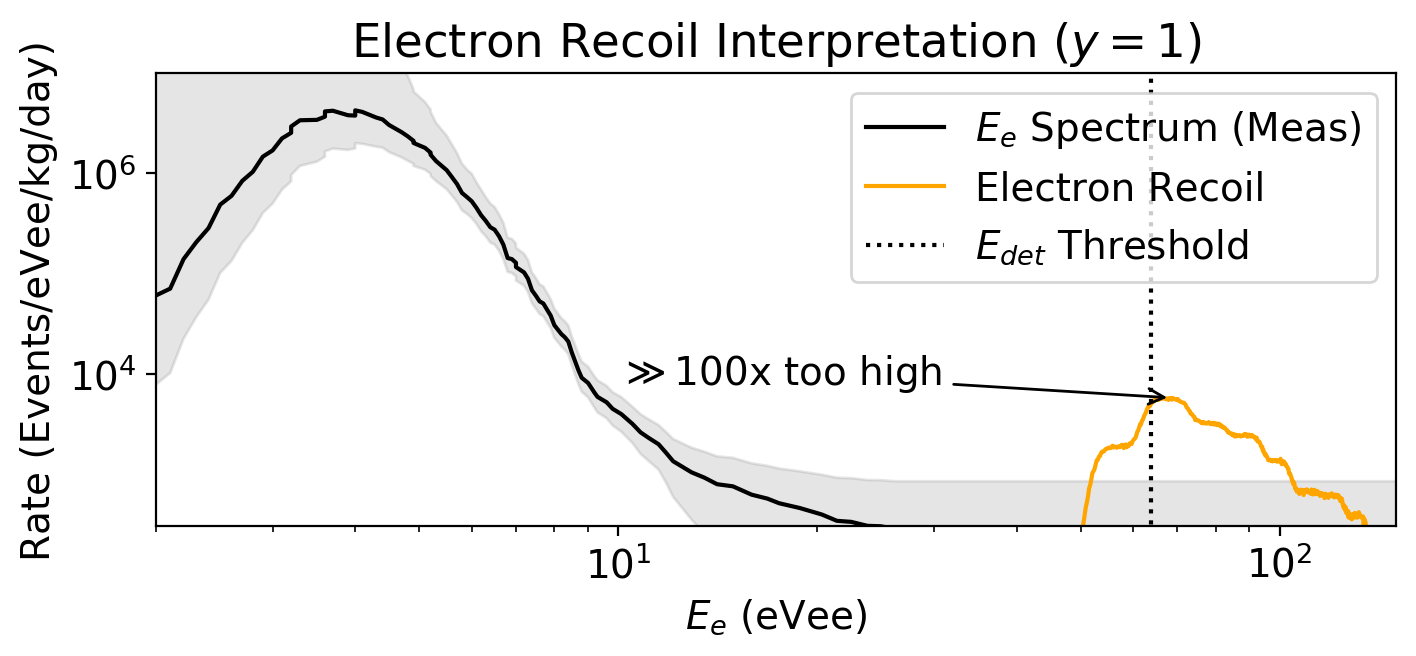} \\
    \includegraphics[width=\textwidth]{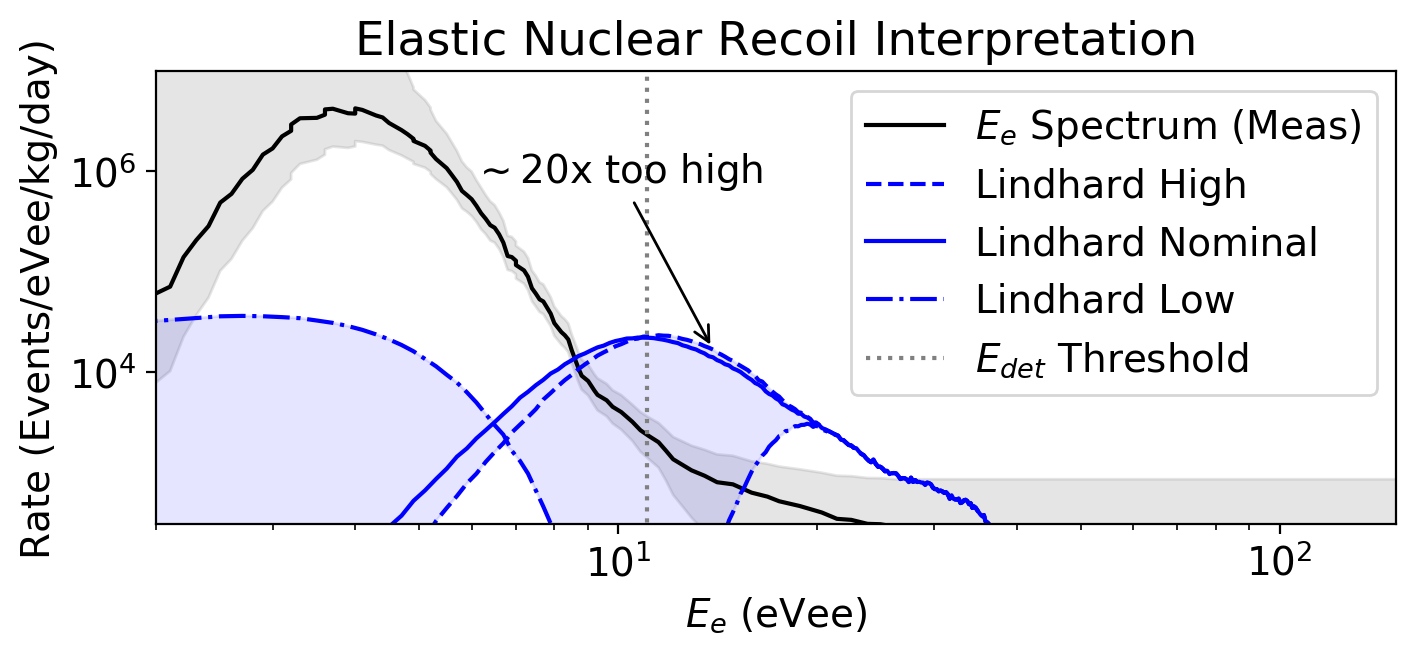}
    \end{minipage}
    \begin{minipage}{0.52\textwidth}
    \includegraphics[width=\textwidth]{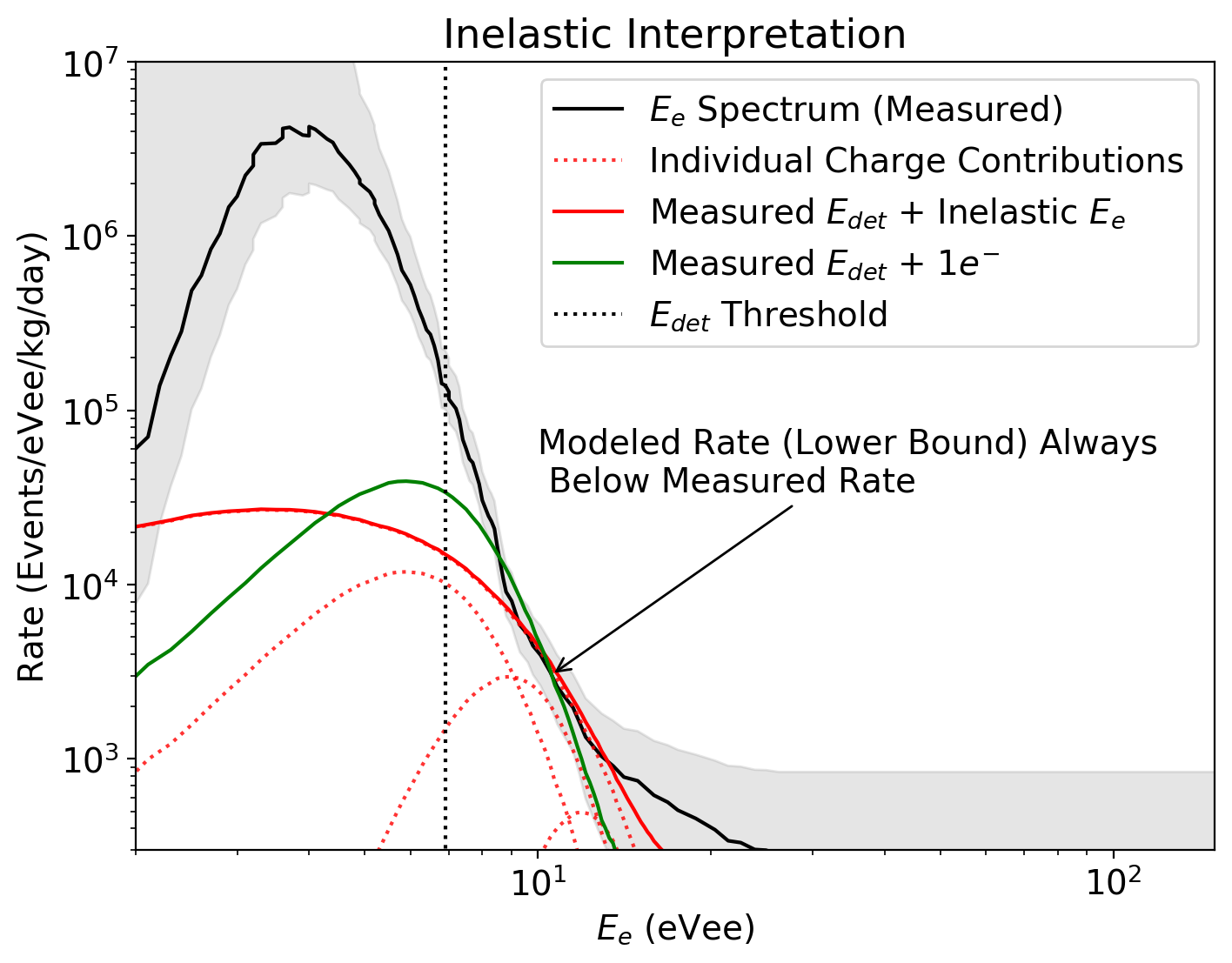}
    \end{minipage}
    \caption{ 
    Comparison of measured ionization ($E_e$) \cite{edelweissHV} and calorimetric ($E_{det}$) \cite{EdelweissWIMP} spectra from EDELWEISS in Ge with yield models (see Appendix~\ref{app:recoil} for further details) for converting the total energy $E_{det}$ into electron-equivalent ioniation energy $E_e$, with uncertainty on the measured $E_e$ spectrum due to statistical error and threshold effects shown in shaded grey.
    \textbf{The $E_{det}$ measurement is a lower bound on the differential rate once converted to $E_e$ by a yield model. For a model to be viable, it must predict an $E_e$ spectrum that is less than or equal to the measured spectrum; the rate will increase as the $E_{det}$ threshold is lowered. If the model does not fit inside of the black curve, then it is an inconsistent interpretation.} Only the inelastic model is able to fit inside of the measured $E_e$ spectrum. \textbf{Top Left:} $E_{det}$ spectrum interpreted as electron-recoil, with $E_{det} = E_{e}$. \textbf{Bottom Left:} $E_{det}$ spectrum interpreted as nuclear recoil (NR) according to the models described in the text and Appendix~\ref{app:recoil}. The blue shaded region around the NR curves illustrates the possible shape variation in the NR models which can be accommodated by physical variation of the yield model. Note that even the ``Low” model over-predicts the $E_e$ spectrum above 20~eVee. \textbf{Right:} Example of a  yield model in which an average of $\lambda_{eh}\sim 0.5$ charges are produced per event regardless of the energy of the event (solid red, with dashed red showing the contribution from each integer number of charges), representative of a signal which produces very few charges independent of event energy. Also shown is a simpler yield model in which exactly 1 charge is produced for every event regardless of event energy (green).
    }
    \label{fig:Rates}
\end{figure*}

At this point in our discussion we therefore make a bold assumption: that all the excesses in Tab.~\ref{tab:eventrates} are caused by a common source.\footnote{Note here that we do not, at this stage, argue that the common source is the same population of dark matter scattering in each detector. Even if dark matter turns out not to be the explanation for these events, the conclusions made here stand independently of the particular source of events.} We justify this assumption based on the charge-readout semiconductor results, arguing at the very least, that interesting new detector physics is being probed by these experiments. If this is the case, then it stands to reason that any other detector should be sensitive to the same rate of these events, and an excess above a modeled background can be interpreted as arising from the same source. The measurement of a statistically significant excess in Ge in both the $E_{det}$ and $E_e$ channels allows us to consider the nature of these events under the assumption of common origin, with the caveat that the location (and thus background environment) of the detector changed between these two runs.

For the last decade, DM experiments have been rejecting irreducible electron recoil backgrounds using the differing yield between nuclear and electronic recoils, often called the quenching factor, utilizing simultaneous measurements of energy in complementary detection channels (see e.g. Refs.~\cite{Barker_2013,Agnese_2017} and Appendix~\ref{app:recoil}). For solid-state experiments, the readout typically comprises both a heat ($E_{det}$) and charge or light ($E_e$) signal. The charge (or light) yield for an event of energy $E_{det}$ is then computed as
$y(E_{det}) = E_e/E_{det}$, 
where $y=1$ is characteristic of an electron recoil event, and $y < 1$, following a measured yield curve \cite{Agnese_2017}, can be used to select the expected nuclear recoil band.

Taking the example of a charge detector, $E_e$ is a derived parameter based on the empirical fact that, on average, one electron-hole pair is produced per $\epsilon_{eh}$ of $E_{det}$ energy.\footnote{$\epsilon_{eh}$ is a measured material property and varies material to material, and is measured such that $E_e$ in different materials for a given calibration source can be plotted on a consistent energy axis.} In other words, an average of
$n_{eh}=E_{det}/\epsilon_{eh}$
electron-hole pairs is produced for such an event, giving the relation $\label{eq:er} E_e = E_{det} = n_{eh}\epsilon_{eh}$ for electron recoil.
While this relation is usually used to convert measured charge to an equivalent energy spectrum, it can also be used to compare measured $E_e$ and $E_{det}$ spectra from the same source of events to determine whether they are consistent with expectations for electron recoils, nuclear recoils, or neither. For further details, see Appendix~\ref{app:recoil}.

The recent release of the high-voltage EDELWEISS DM search \cite{edelweissHV} is thus the most significant development to date because, taken with the previously published $E_{det}$ spectrum from a similar detector, it is the first dataset for which we can compare the two spectra directly in a single material to determine a likely origin. This type of detector actually measures a combination of $E_e$ and $E_{det}$ as we have defined them, producing an $E_e$ measurement according to
\begin{equation}
    E_e = E_{det}\left[ y(E_{det})+\frac{\epsilon_{eh}}{e\cdot V_{det}}\right],
\end{equation}
where $V_{det}$ is the detector operating voltage and $e$ is the electron charge. This reduces to our definition of $E_e$ only in the limit $V_{det}\rightarrow\infty$; the data considered here were taken at 78~V. This gives an additional correction term of $\epsilon_{eh}/e V_{det} \sim 3.8\times 10^{-2}$ for $\epsilon_{eh}=3.0$~eV \cite{Agnese_2017}.

Figure~\ref{fig:Rates} shows these spectra under three scenarios for the origin of the $E_{det}$ spectrum, assuming it originates from a single type of event:
\begin{enumerate}
    \item{\bf Electron Recoil Interpretation:} The events are electron recoils, with $y = 1$ (Figure~\ref{fig:Rates}, top left). This is clearly inconsistent because the black and orange curves are markedly different, and electron recoils are strongly ruled out. 
    \item{\bf Elastic Nuclear Recoil Interpretation:} The events are nuclear recoils, and the yield follows the measured Ge nuclear recoil yield model \cite{Agnese_2017} with varying low-energy behavior (Figure~\ref{fig:Rates}, bottom left). We consider an extrapolation of measured yield to the bandgap energy (Nominal), a yield constant below 100~eV (High), or a yield that drops discontinuously to 0 at 100~eV (Low). All are clearly also inconsistent with the measured $E_e$ spectrum.
    \item{\bf Inelastic Interpretation:} Finally, we consider a maximally inelastic yield, in which \textit{every} event produces a charge yield independent of the recoil energy, such that
    \begin{equation}
        \langle E_e \rangle = \epsilon_{eh}  \left(     \lambda_{eh}+\frac{E_{det}}{e\cdot V_{det}}   \right), 
    \end{equation}
    where $\lambda_{eh}$ is the mean number of electron-hole pairs produced by the event (Figure~\ref{fig:Rates}, right). Unlike the previous two cases, this matches the observed spectrum remarkably well both in signal shape and event rate for $\lambda_{eh}\leq0.5$ (no relative scaling is done to force the rate to match), suggesting an inelastic interpretation is allowed for the measured $E_e$ spectrum, in contrast to the two standard scenarios.
\end{enumerate}

Based on this simple analysis, we conclude that an interpretation of the EDELWEISS events based on standard elastic NR or ER models is inconsistent, and that the most likely interpretation of these events is an inelastic interaction, with a yield curve that \textit{increases} with lower event energy. If this is the case, it also helps reconcile the event rates in well-ordered crystals (which see a rate of $\mathcal{O}$(10 Hz/kg)) compared to liquids or amorphous solids, which observe much smaller event rates. An inelastic interaction will be largely driven by condensed matter properties unrelated to the nuclear mass or electron density we use to relate different targets to each other under standard elastic assumptions.


This observation therefore rules out ``standard'' backgrounds caused by known low-energy interactions of photons, charged particles, and neutrons. It does not preclude the aforementioned crystal cracking events, which would not intrinsically produce light or charge. However, we note at this point that, if crystal cracking events were truly causing the $E_e$ background, one would expect there to be a dependence on applied pressure, temperature, and operating history. 
We therefore either have to accept a crystal cracking rate determined \textit{only} by material, or ask what other physical process might lead to a consistent rate with an inelastic-like charge yield.



\section{Plasmon Interpretation}
\label{sec:Plasmon}
\subsection{Plasmon Properties}
Without committing to a particular source of signal events yet, we postulate that the nature of the observed excitations in low-threshold silicon, germanium, and sapphire detectors is the plasmon.\footnote{In this work ``plasmon'' will only refer to a bulk plasmon, in contrast with surface plasmons which are qualitatively different phenomena.} The plasmon model, presented here, is consistent with the maximally inelastic yield discussed in Section~\ref{sec:SignalOrigin} given the known properties of plasmons. In this section we briefly review the properties of plasmons relevant for our analysis; see Appendix~\ref{app:Plasmon} for more details.

A plasmon is a long-wavelength collective excitation of charges in a lattice which carries energy near the classical plasma frequency,
\be
E_p \simeq \sqrt{\frac{4\pi \alpha n_e}{m_e}},
\ee
where $\alpha$ is the fine-structure constant and $n_e$ is the electron number density; in a semiconductor, $n_e$ is to be interpreted as the density of valence electrons.\footnote{Plasmons can also appear in ``metamaterials,'' where $n_e$ is interpreted as the average electron density in a heterostructure averaged over large distances; see \cite{Lawson:2019brd} for a proposal to use these plasmons to detect axion DM.} Since most solid-state systems have roughly the same number density, with interatomic spacing of a few Angstroms, $E_p \sim \mathcal{O}(10-100) \ \eV$ across essentially all materials (see Tab.~\ref{tab:plasmons}). In particular, bulk plasmons exist and have been observed in silicon, germanium, and sapphire. The long-wavelength nature of the plasmon is reflected in a momentum cutoff 
\be
\label{eq:qc}
q_c \sim \frac{2\pi}{a} \sim 5 \ \keV~,
\ee
where $a$ is the lattice spacing. If a plasmon carries $q > q_c$, it represents a charge oscillation localized to within a single lattice site, and the plasmon will decay very rapidly into a single electron-hole pair in a process known as Landau damping \cite{raether2006excitation}. Note that the creation of such a short-range plasmon is inconsistent with the analysis of the Ge spectra in Sec.~\ref{sec:SignalOrigin} above, which suggests that the plasmon should have a dominant decay channel into phonons only. Thus we will focus exclusively on excitation of long-range ($q<q_c$) plasmons.

\begin{table}[b]
    \centering
    \begin{tabular}{|c|c|c|}
        \hline
        Material & Plasmon Energy $E_p$ (eV) & Width $\Gamma$ (eV) \\
        \hline
        Si & 16.6 & 3.25 \\
        Ge & 16.1 & 3.65 \\
        Al$_2$O$_3$ & 24.0 \cite{PhysRevB.32.1237} & $\sim 5$\\
        GaAs & 16.0 & 4.0 \\
        Xe (Solid) & 14--15 \cite{Nuttall_1975} & $\sim 4$\\
        Ar (Solid) & 19--21 \cite{Nuttall_1975} & $\sim 5$\\
        \hline
        CaWO$_4$ &  \multicolumn{2}{c|}{Unknown} \\
        \hline
    \end{tabular}
    \caption{Plasmon energies in various materials. Crystal values taken from Ref.~\cite{kundmann1988study} unless otherwise referenced. We were unable to find measurements of plasmon features in CaWO$_4$, and expect that it has a much weaker plasmon resonance than the other crystals considered here. It is significant to note that the solid forms of the noble elements show strong resonance features; the liquid forms do not.}
    \label{tab:plasmons}
\end{table}


The plasmon is most easily observed in electron energy-loss spectroscopy (EELS), where fast ($\sim 100$ keV) electrons impinging on a material have a high probability of depositing energy $E_p$. This probability is only weakly dependent on the incident electron energy $E_0$, scaling as $\log(E_0)/\sqrt{E_0}$ (see Appendix~\ref{app:Plasmon}), and is independent of the target material except for the core electron contribution to the dielectric constant. At the same time, the probe must be fast in order to deposit a small amount of momentum for a given energy $E_p$. In other words, probes with sufficient energy $E_0 \gg E_p$ and sufficient velocity will strongly prefer to deposit energy $E_p$, regardless of their initial energy, at similar rates across diverse materials. This behavior is typical of other resonances encountered in nuclear physics or electrical engineering; in a sense, the plasmon acts as a band-pass filter for $E_{det}$.

The lineshape of the plasmon near the peak is well described by a Lorentzian \cite{kundmann1988study}, where the finite width $\Gamma$ parameterizes the decay of the plasmon into phonons and/or electron/hole pairs, which are the long-lived excitations in the detector. We note that the plasmon is inherently a many-body excitation, and cannot be described in terms of non-interacting single-particle states, such as band structure wavefunctions derived using density functional theory \cite{RevModPhys.28.184}. Moreover, typical values of $\Gamma/E_p$ for semiconductors are of order $\sim$ 0.2 \cite{kundmann1988study}, which is comparable to $\Gamma/M$ for the $\rho$ meson and larger than $\Gamma/M$ for most other strongly-decaying hadronic resonances, and suggests that the couplings which govern plasmon decay are large or even nonperturbative. The simple yield model for the Ge spectra suggests that the plasmon must have a $\sim$50\% branching fraction to phonons only. To our knowledge, the branching fractions of the plasmon to phonons or electron/hole pairs is unknown, but in principle these could be determined from a suitably modified EELS experiment with both calorimetric and charge readout. 



Based on this interpretation, assuming some incident flux of particles is dominantly exciting the plasmon over other elastic or inelastic excitations, detectors with $E_{det}$ thresholds approaching $E_p$ from above should see a sharp rise in events as the threshold is lowered; this qualitatively explains the results from the silicon, germanium, and sapphire experiments, as well as the null results from previous experiments with thresholds well above $E_p$. Moreover, the plasmon in germanium has a significant high-energy tail and double-peaked structure resulting from contributions from the $3d$ shell \cite{kundmann1988study}, further explaining the onset of events in EDELWEISS despite a threshold of $60 \ \eV \sim 4E_p$. By contrast, the plasmon in silicon lacks a corresponding tail, explaining the lack of a signal excess in higher-threshold analyses of DAMIC \cite{PhysRevD.94.082006} and CDMSlite \cite{Agnese_2019} data. Furthermore, materials without long-range order such as liquid xenon and, to a lesser extent, CaWO$_4$ do not have a pronounced plasmon peak, consistent with the lower event rates from XENON10 and CRESST.

\subsection{Plasmons from Known Particles?}

An interpretation of the plasmon excitation as sourced by SM particles or fields is extremely difficult.
\begin{itemize}
    \item{\bf Photons and electromagnetic fields:} Transverse UV and soft X-ray photons cannot source the longitudinal plasmon oscillation, and static electric fields cannot source oscillating charges.
    
    \item {\bf Charged SM Particles:}
    The inelastic mean free path for charged particles such as electrons or muons, or for x-rays, is on the order of tens of nm, so these particles would be expected to undergo multiple scattering and deposit many multiples of $E_p$ as they traversed a detector (all of which are much thicker than nm for the experiments we consider), which would lead to many events above threshold contrary to what was observed. A single energy deposit under 100 eV is only consistent with a particle of mean free path much larger than the detector thickness; if charged, this particle would have to have electric charge much less than $e$.
    
    \item {\bf Neutrons:} In principle, it is possible that hard scattering events induced by neutrons may create secondary plasmon excitations; indeed, we speculate on this possibility in Sec.~\ref{sec:Scenario1} below in the context of hard DM-nucleus scattering. However, one would have to explain why the neutron flux is the same at all the semiconductor experiments listed in Table I regardless of the shielding, detector environment, detector construction, and exposure.
    
    \item{\bf Neutrinos: }  Astrophysical neutrinos can, in principle, undergo neutral-current scattering with a seminconductor nucleus whose recoil excites a plasmon independently of detector overburden. However, the known solar and atmospheric fluxes (assuming SM weak interactions) cannot account for rates of the observed magnitude \cite{Harnik:2012ni}. We can conservatively estimate this contribution by considering solar $pp$ neutrinos whose peak flux is $\sim  10^{11}$ cm$^{-2}$ s$^{-1}$ near their kinematic endpoint at $E_\nu \sim 400$ keV \cite{Bellerive:2003rj}. The total coherent neutrino-nucleus scattering
    cross section on Ge targets is approximately $\sigma_{\nu-\rm Ge}\sim 10^{-42}$ cm$^2 (E_\nu/400 \, \rm keV)^2$ \cite{Scholz:2017ldm},
    so the total event rate from $pp$ neutrinos is roughly $\sim 10^{-6}$ Hz/kg, which is many orders of magnitude below
    the low-threshold excess rates observed in semiconductors; other populations of solar or atmospheric neutrinos have considerably lower fluxes.
    Although it may be possible for an unknown population of very low-energy  neutrinos to excite plasmons through non-standard (larger than electroweak) interactions,  
exploring this scenario is beyond the scope of the present work. 
\end{itemize}

We conclude that none of these options offers a satisfactory explanation for the observed excesses.

\section{Dark Matter Scenarios For Plasmon Excitation}
\label{sec:Models}

 \begin{figure}
     \includegraphics[width=0.45\textwidth]{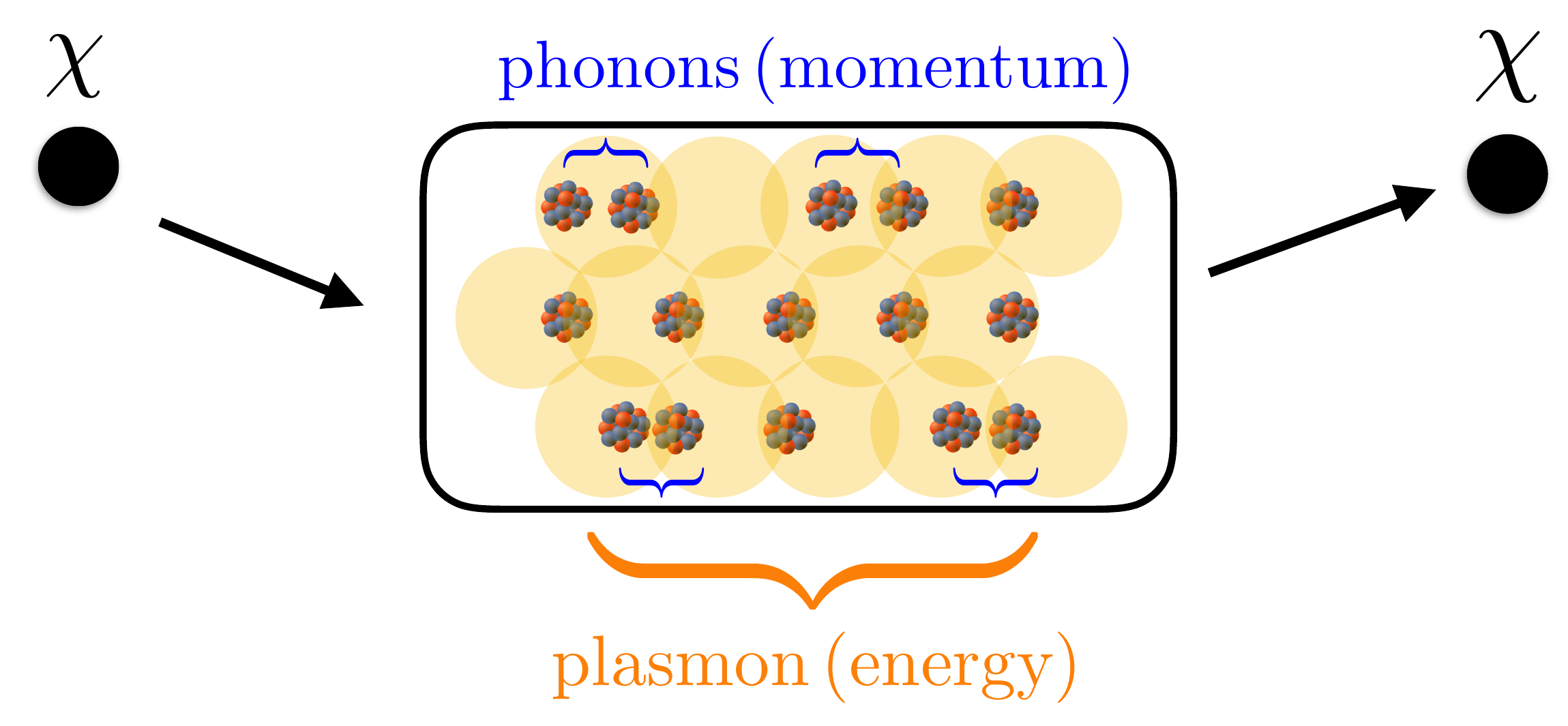}
          
     \caption{Cartoon of indirect plasmon excitation through a hard scattering event, where the imparted momentum is dominantly carried by multiple phonons, while the imparted energy is carried by the low-momentum plasmon.}
     \label{fig:Cartoon}
 \end{figure}

\begin{figure*}
    \hspace{-0.5cm} \includegraphics[width=0.5\textwidth]{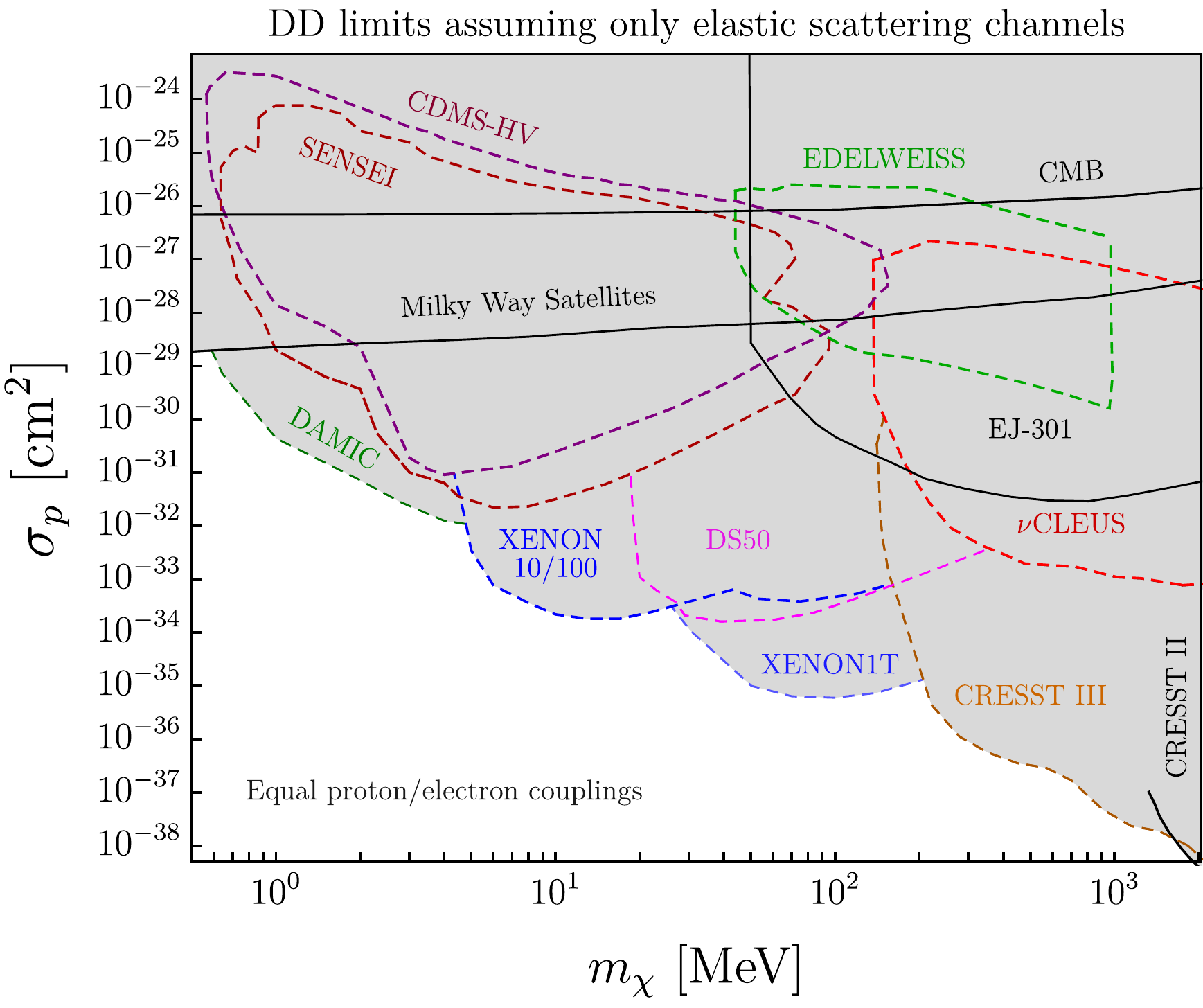}~~
     \includegraphics[width=0.5\textwidth]{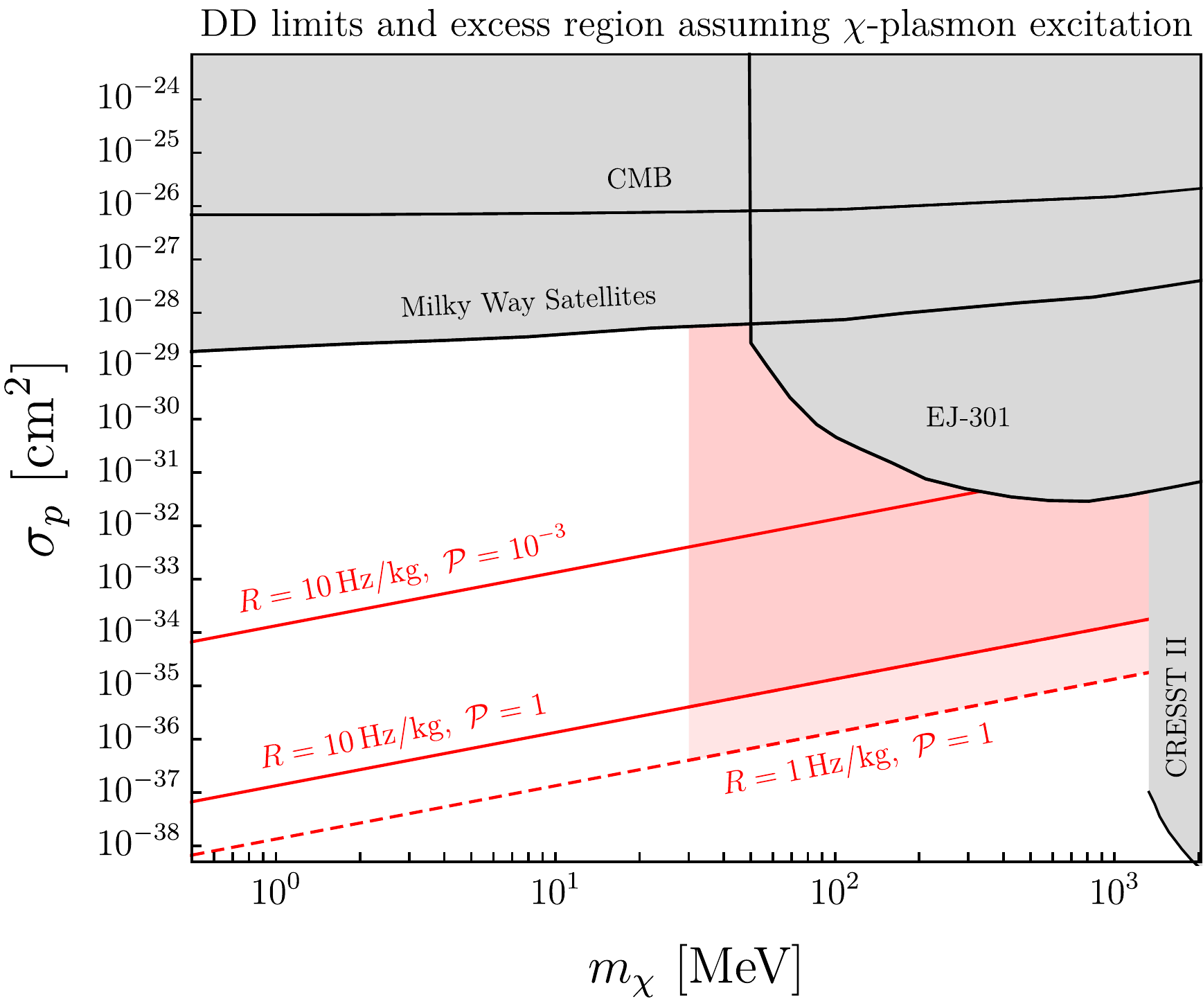}
     \caption{{\bf Left:} 
     If DM-SM scattering is assumed to involve elastic interactions with detector particles, there are many reported bounds on $\sigma_p$ assuming equal DM couplings to electrons and protons.
     Note that several of the bounds in this plot are based on translating electron recoil searches according to $\sigma_p = \sigma_e \mu^2_{\chi p}/\mu^2_{\chi e}$ where $\mu_{ij}$ is the $ij$ reduced mass. However {\it every} experiment shown in dashed contours observes an excess of events as shown in Table \ref{tab:eventrates} (see also \cite{Emken:2018run,Emken:2019tni} for discussions of the upper part of the contours). These excesses are not currently reported as DM signals because the spectral shape for elastic DM scattering does not provide a good fit to these data (see the left panel of Fig.~\ref{fig:Rates} for examples of such shape mismatch). 
      Consequently, these results are reported as limits, not as evidence of a DM signal.
     {\bf Right:}
     Favored parameter space for which an DM-proton interaction with an a secondary plasmon excitation probability $\cal P$ 
     can accommodate excess event rates in a Ge target on the order of $1-10$ Hz/kg with $E_{det}$ up to 100 eV (shaded pink and red).
      Note that the CRESST-II \cite{Angloher_2016} bounds are based
     on a nuclear recoil search, so these constraints still apply; the additional parameter space covered by CRESST-III is also the region in which an excess was seen. Although we are arguing that, in this scenario, the constraints from the left panel do not apply straightforwardly  due to their observed excesses, these bounds are also model-dependent and inapplicable if the DM interacts only with nucleons since the low-threshold bounds assume electron recoil signals. Furthermore, a leptobhobic DM-nucleus interaction could excite plasmons without ever directly inducing electron recoils, so many of the dashed
     regions in the left panel may not even apply in principle. The one exception is the limit from EJ-301~\cite{collar_simp} which does observe an excess rate but sets conservative limits based only on the total single-photoelectron rate rather than a fit to an expected spectrum. 
     Also shown are exclusions from CMB scattering \cite{Xu:2018efh}
     and Milky Way satellites \cite{Nadler:2019zrb}.}
     \label{fig:contact_interaction}
 \end{figure*}

\begin{figure}
    \hspace{-0.5cm}  \includegraphics[width=0.5\textwidth]{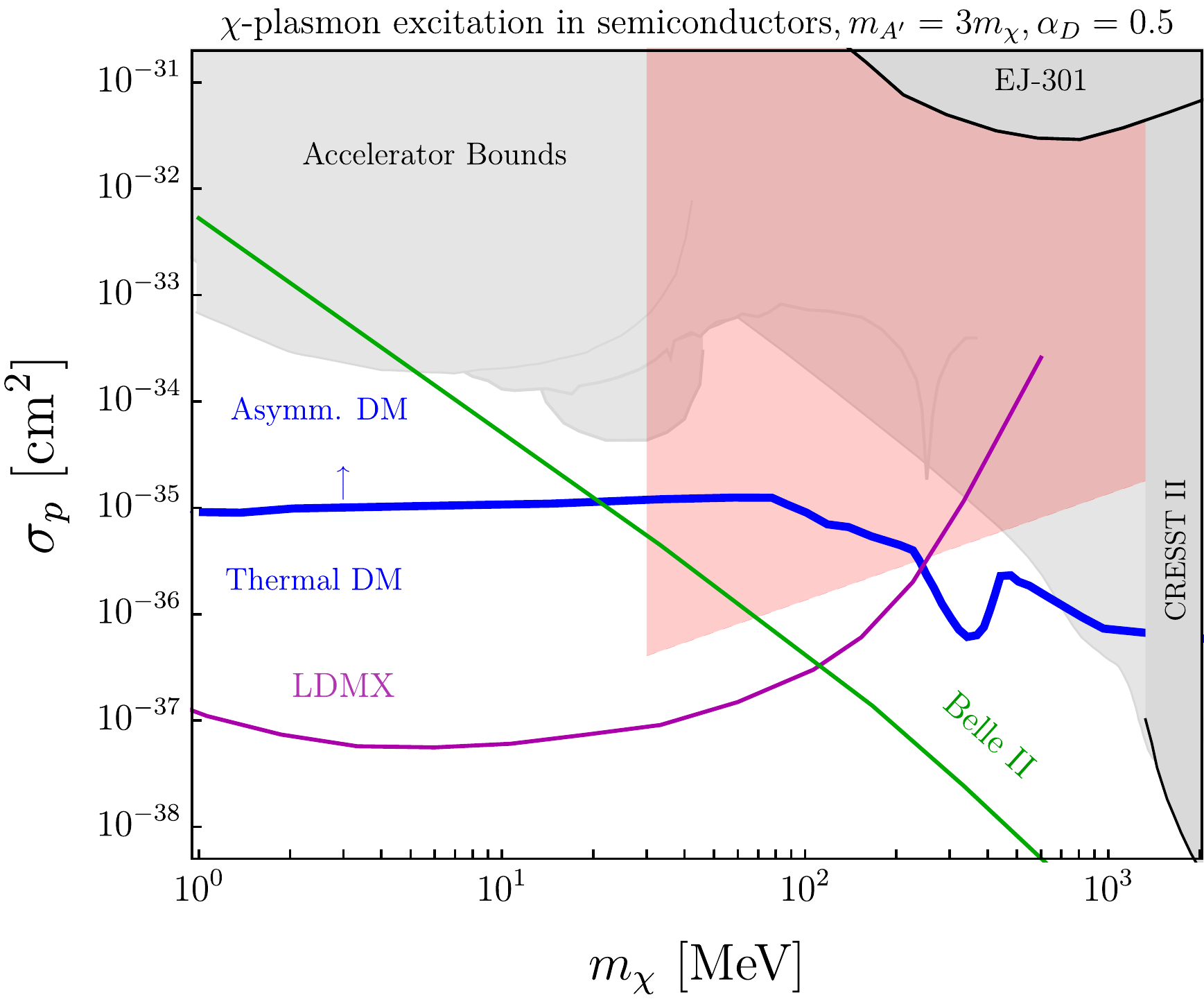}
     \caption{ Same red shaded favored parameter space as Fig. \ref{fig:contact_interaction} (right panel) interpreted in the context of dark photon mediated scattering (see Scenario 1, Sec.~\ref{sec:Scenario1}). The blue curve represents the parameter space for which direct $\chi\chi \to$ SM SM annihilation  
     accounts for the full DM abundance; parameter space above this curve can be viably accommodated if the DM population is particle-antiparticle asymmetric.
     The additional gray shaded region below the bounds shown in Fig.~\ref{fig:contact_interaction} (right panel) represents accelerator based constraints for a representative choice of parameters $m_{A^\prime} = 3 m_\chi$ and $\alpha_D = 0.5$. This envelope contains bounds from the LSND \cite{deNiverville:2011it}, MiniBooNE \cite{Aguilar-Arevalo:2018wea}, and E137 beam dumps \cite{Batell:2014mga,PhysRevLett.121.041802}, BaBar \cite{Essig:2013vha}, and NA64 \cite{NA64:2019imj}. Also shown are projections for the LDMX 
     missing momentum experiment \cite{Berlin:2018bsc} and the $B$-factory Belle-II \cite{Kou:2018nap}.
     As in Fig \ref{fig:contact_interaction} (right), the pink shaded region is compatible with a 10 Hz/kg plasmon excitation rate in Ge for a virialized DM halo population; the region to the left can also accommodate this rate if a DM subcomponent is faster than the Galactic escape velocity.
     }
     \label{fig:contact_interaction_darkphoton}
 \end{figure}

Having excluded the possibility that the plasmon could arise from SM particles, we now make a further leap and consider the hypothesis that DM could account for these plasmon excitations.  If a DM particle with mass $m_\chi$ and incident velocity $\vecv$ deposits energy $E$ and momentum $\vecq$ in a detector, energy conservation requires
\be
E = \vecq \cdot \vecv - \frac{q^2}{2m_\chi},
\ee
which implies
\be
\label{eq:qcond}
q \geq \frac{E}{v},
\ee
which is saturated in the limit of forward scattering and $m \to \infty$. Taking $E = E_p = 16 \ \eV$ for the typical plasmon energy in Ge, we find that to excite the plasmon directly (i.e. $q < q_c$, see Eq.~(\ref{eq:qc})) we must have 
\be
v \gtrsim 10^{-2} \ {(\rm direct \ plasmon \ excitation)}.
\ee
Since this exceeds Galactic escape velocity in the Earth frame \cite{Evans:2018bqy}, gravitationally-bound DM with $v \sim 10^{-3}$ cannot {\it directly} excite a long-range plasmon. However, we identify two qualitatively distinct mechanisms by which DM (or a sub-component) can still account for the observed excesses: 

\begin{itemize}
    \item {\bf Scenario 1, Secondary Plasmon:} 
    In analogy with the Migdal effect \cite{Migdal1939,Ibe:2017yqa}, if halo DM with the 
    standard Maxwellian velocity distribution peaked at $v\sim 10^{-3}$ first scatters off a target nucleus, the interaction can transfer a majority of the momentum to phonons, while imparting most of the deposited energy to the plasmon which carries $q<q_c$ (see Fig.~\ref{fig:Cartoon}).\footnote{Note that the scale of the momentum transfers we will consider, $\mathcal{O}(15 \ \keV)$ from Eq.~(\ref{eq:qcond}), is precisely in the regime between single-phonon excitation and direct nuclear scattering, where direct multi-phonon production is expected to dominate \cite{Trickle:2019nya}. Indeed, the displacement energy of bulk Ge is 10--50 eV \cite{jiang2018theoretical}, so below this energy, an elastic nuclear recoil is not even an on-shell state, and the non-electronic energy must appear in the form of phonons -- see Appendix~\ref{app:recoil}.}  The plasmon can then decay to phonons and electron/hole pairs. In this scenario, the signal rates scale as $Z^2$
    where $Z$ is the atomic number of the target material.
    
    \item{\bf Scenario 2, Fast DM Sub-Component: }
    Although the majority of halo DM in our Galaxy must satisfy $v \lesssim 10^{-3}$ to
    account for observed rotation curves, it is possible
    that a small fraction $f \ll 1$ of the local DM density is accelerated to speeds $v \gtrsim 10^{-2}$ above Galactic escape velocity (e.g by solar reflection \cite{Emken:2017hnp,An:2017ojc}).\footnote{Other possibilities for achieving a fast sub-component of dark sector particles include boosted DM \cite{Agashe:2014yua, Necib:2016aez, Berger:2019ttc}, cosmic ray up-scattering \cite{Bringmann:2018cvk,Cappiello:2019qsw,Dent:2019krz,Krnjaic:2019dzc}, direct production in supernovae \cite{Chang:2018rso,DeRocco:2019jti}, and acceleration from supernova remnants \cite{Li:2020wyl}.}  Unlike in Scenario 1 above, here the rate scales inversely with the target's mass density and is independent of $Z$ since the plasmon is excited directly without the DM having to first undergo nuclear scattering. 
\end{itemize}
These scenarios are complementary: Scenario 1 requires no non-standard DM ingredients but features large theoretical uncertainty in the plasmon-phonon coupling; by contrast, Scenario 2 has no theoretical uncertainty in the direct plasmon excitation probability, which is in one-to-one correspondence with an EELS measurement, but requires an explanation for the fast DM sub-component. In both scenarios, a plasmon with a large branching ratio to phonons only can accommodate the spectral shape of the excess and
match the total observed rate in the EDELWEISS 78 V run for 2 or more charges, $\sim$ 20 Hz/kg \cite{edelweissHV}.

Theoretically, both of these scenarios  can be realized within a standard framework for DM below the GeV scale. Let $\chi$ be a DM candidate particle of mass $m_\chi$ coupled to a new spin-1 $U(1)$ gauge boson $A'$, which kinetically mixes with the SM photon. Here
$\chi$ can be a scalar or a fermion and such an interaction has long been a standard benchmark for sub-GeV DM studies \cite{Essig:2013lka,Alexander:2016aln,Battaglieri:2017aum}. In the mass eigenbasis, the Lagrangian for this model can be written
\be
\mathcal{L} \supset - \frac{m_{A'}^2}{2} A'_{\mu}A'^{\mu} + A'_\mu( \kappa e J_{\rm EM}^\mu +  g_D  J^\mu_D) ,
\ee
where $J_{\rm EM}$ is the SM electromagnetic current, $\kappa \ll 1$ is a small kinetic mixing parameter, $g_D$ is the DM-$A^\prime$ coupling constant and $J^\mu_D$ is the DM current
\be
J_D^\mu = \begin{cases}
i(\chi^* \partial^\mu \chi - \chi \partial^\mu \chi^*)~&\text{Scalar}
\\
\bar \chi \gamma^\mu \chi ~&\text{Fermion},
\end{cases}
\ee
which are analogous to scalar and fermionic versions of
``dark electromagnetism" with a massive dark photon. 

In the limit where the dark photon is massless, $m_{A'} \to 0$, the DM effectively acquires an electric millicharge $\kappa g_D$; this interpretation holds as long as $m_{A'} \ll q$ where $q$ is the typical momentum transfer in the process under consideration. In the opposite limit, where $m_{A'} \gg q$, the DM effectively has contact interactions with charged particles, including electrons and nuclei.

\subsection{Scenario 1: Secondary  Plasmon Excitation through Hard Inelastic Scattering}
\label{sec:Scenario1}
One way to interpret the origin of this plasmon resonance signal is through the inelastic nuclear scattering of 100~MeV-scale DM through a contact interaction ($m_{A'} \gg q$). 
This is similar to recent calculations of the Migdal effect~\cite{Ibe:2017yqa}, except that the existing literature presenting the formalism for the Migdal effect relies on an isolated atom approximation and cannot be reliably extended to semiconductors.\footnote{Ref.~\cite{Essig:2019xkx} made strides toward addressing this problem, though their analysis was still restricted to non-interacting single-particle wavefunctions, which cannot describe the plasmon.}

In light of this uncertainty, we factorize the DM-induced ionization rate $R$ in a semiconductor into a spin-averaged single-proton cross section
\be
\sigma_p = \frac{16\pi \kappa^2 \alpha \alpha_D \mu_{\chi p}^2}{m_{A'}^4},
\ee 
where $\alpha_D \equiv g_D^2/4\pi$,
times an energy/momentum-averaged plasmon excitation probability $\mathcal{P} \leq 1$ per individual nuclear scatter, such that the total plasmon excitation rate is
\be
R = N_T \mathcal{P} \frac{\rho_\chi}{m_\chi} Z^2 \sigma_p v,
\ee
where $\rho_{\chi} = 0.4 \, {\rm GeV\, cm^{-3}}$ is the local DM density \cite{Bovy:2012tw}, $N_T$ the number of detector targets, $v$ is the DM velocity, and $Z$ is its atomic number. Note that $\mathcal{P}$ is a property of the detector material, and would be expected to vary somewhat between Si and Ge, for example, but should be similar across all single-crystal detectors of the same material. Our assumption in this scenario is that, despite the fact that the DM transfers momentum $q \gg q_c$ to the material, a long-wavelength plasmon with $q < q_c$ is excited, with phonons or Umklapp processes absorbing the remainder of the momentum. The factor of $Z^2$ in the rate arises because the maximum momentum transfer for sub-GeV DM is $q_{\rm max} = 2 m_\chi v \sim 2 \ \MeV$, which is not large enough to probe the nuclear structure, so the interaction is coherent over all the protons in the nucleus (i.e. the nuclear form factor is unity). For Ge, we find
 \begin{equation}
\!\!\! R \sim   \frac{10 \ \rm Hz}{\rm kg}   \left( \frac{\cal P}{0.1} \right)
\left( \frac{\sigma_p }{ 7 \times 10^{-35} \, \rm cm^2 }\right)\! \left( \frac{50 \ \MeV}{m_\chi} \right).
 \end{equation}  
In Fig.~\ref{fig:contact_interaction} (left), we show the constraints on the relevant parameter space when limits from the various experiments shown in Tab.~\ref{tab:eventrates} are interpreted in terms of traditional electron recoil or nuclear recoil models. These bounds do not apply in the indirect plasmon excitation model we a considering here; in Fig.~\ref{fig:contact_interaction} (right) we show only the direct nuclear recoil bounds from CRESST-II and EJ-301 which survive. Indeed, we are proposing that many of the ``bounds,'' which correspond to actual low-energy excesses in the data, are in fact signals when interpreted in the plasmon model. We shade in red the region where DM at the Galactic escape velocity has kinetic energy $\frac{1}{2}m_\chi v^2 > 100 \ \eV$ in order for gravitationally-bound DM to explain the observed $E_{det}$ spectrum in Ge. In Fig.~\ref{fig:contact_interaction_darkphoton} we show the same parameter space including a variety of accelerator-based bounds and projections for future searches. If this scenario is correct, various fixed-target and $B$-factory follow-up measurements will be sensitive to the full parameter space responsible for the low threshold direct detection excesses.

For each value of $m_\chi$, there is a specific value of $\sigma_p$ which would generate the observed cosmological DM abundance through thermal freeze-out, or in the case of asymmetric dark matter, provides a lower bound on the cross section required to annihilate away the symmetric component. This line is shown in blue in Fig.~\ref{fig:contact_interaction_darkphoton} \cite{Izaguirre:2015yja,Berlin:2018bsc}. We see that for $\mathcal{P}$ of order 1, and a DM mass of $30-200$ MeV, the multiple excesses described in Table \ref{tab:eventrates} can be consistently explained with a dark photon interaction that also sets the relic density to SM particles in the early universe.

\begin{figure*}
     \includegraphics[width=0.45\textwidth]{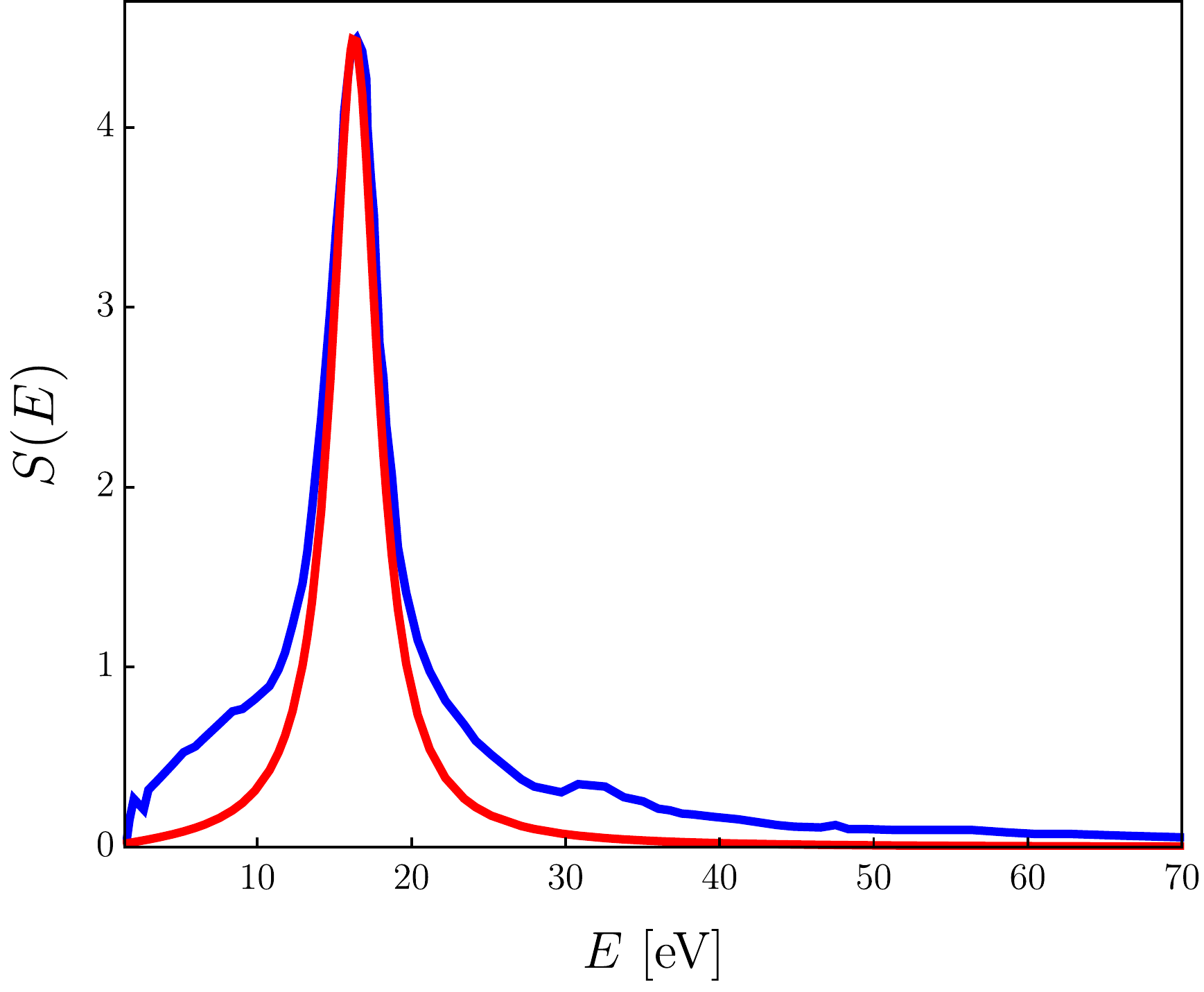}~~
     \includegraphics[width=0.475\textwidth]{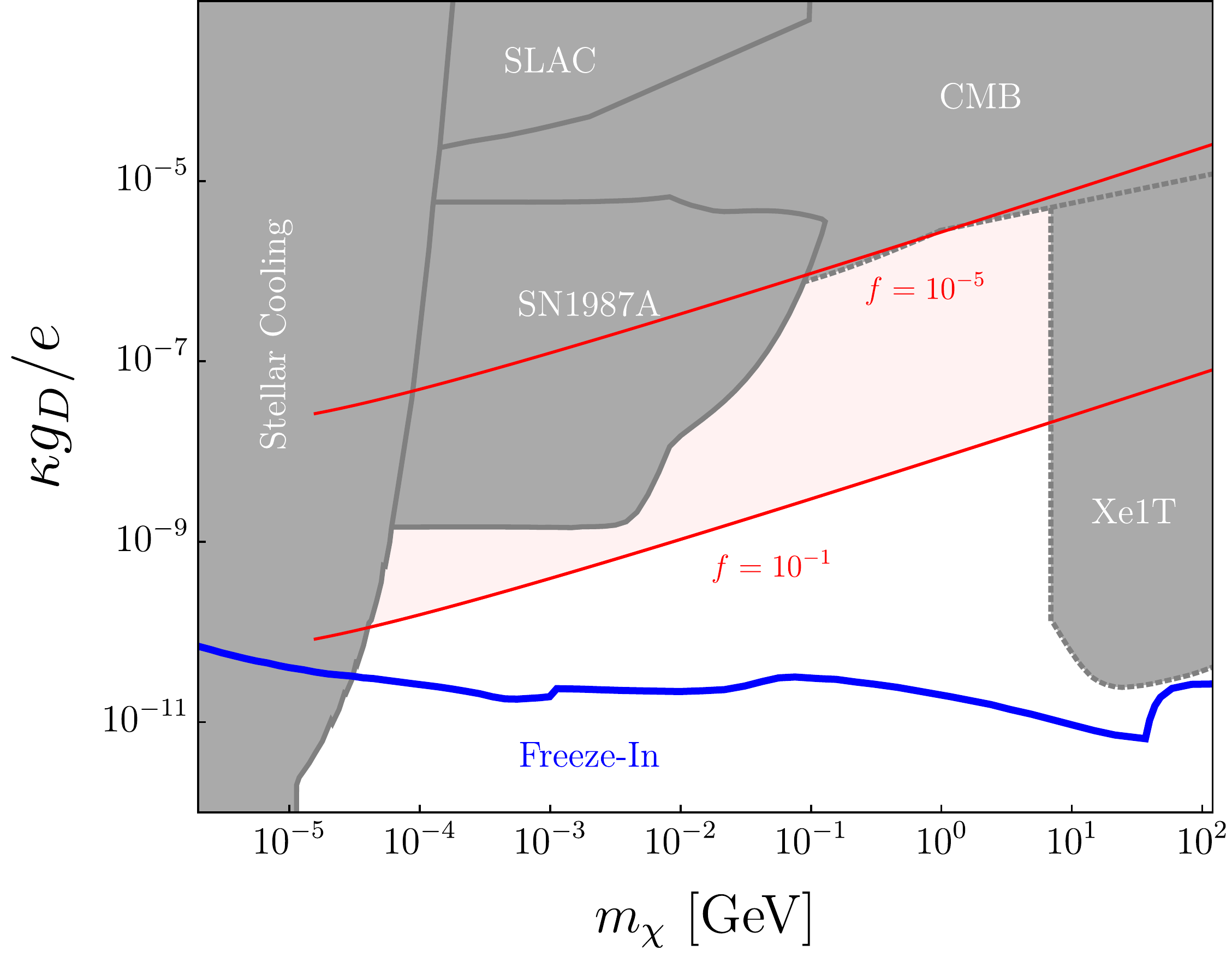}
          
     \caption{ {\bf Left: } Lineshape $S(E)$ of the germanium plasmon from EELS measurements \cite{kundmann1988study} (blue), normalized to match the peak of
     the best-fit Fr\"{o}hlich model (red) described in Appendix~\ref{app:Plasmon}. {\bf Right:} Parameter space for a subcomponent of semi-relativistic DM interacting through a light mediator, expressed in terms of the effective millicharge $\kappa g_D/e$. The shaded pink region shows the preferred parameter space for the 20 Hz/kg plasmon signal in EDELWEISS, assuming a fraction $f$ of the local DM density has velocity $v = 0.1$, for $f$ ranging from $10^{-5}$ to $10^{-1}$. Exclusions from SN1987A~\cite{Chang:2018rso}, CMB~\cite{Dvorkin:2019zdi},
     SLAC \cite{Diamond:2013oda}, 
     XENON1T~\cite{xenon1t}, and stellar cooling \cite{Davidson:2000hf} are shown in grey. }
     \label{fig:light_mediator}
 \end{figure*}

Although a large value of $\mathcal{P} \sim {\cal O}(1)$ is somewhat surprising, it is not unreasonable given that the consistency of this explanation requires a large branching fraction of the plasmon to phonons. As we describe in Appendix~\ref{app:Plasmon}, the plasmon is best understood as a nonperturbative effect with relative width even exceeding that of QCD resonances, so large couplings are expected. Intriguingly, the interaction we are proposing is maximally coherent in the sense that it benefits from coherence over the nucleus in the hard scattering event, followed by coherence over the the lattice sites in the excitation of the plasmon, such that there is no suppression by factors of momentum anywhere in the rate. The conventional wisdom is that such coherent events are always suppressed by either a small momentum or small phase space \cite{Akhmedov:2018wlf}, but such arguments typically involve a single-particle picture, and it is plausible that this intuition is modified by many-body effects in condensed matter systems. To our knowledge, such a secondary excitation process has not been previously considered in the condensed matter literature, though the plasmon-phonon coupling has been computed for polar semiconductors \cite{10.1143/PTP.57.97}. We make some suggestions in Sec.~\ref{sec:Conclusions} for neutron experiments which may confirm this effect. Regardless, the near-perfect match between the observed rate and the expected cross section for thermally-produced DM suggests that this process should be taken seriously as a signal candidate. Indeed, the rates studied here should be considered the {\it largest} rates able to accommodate this well-motivated model, as some portion of the total integrated rate is expected from true dark rate backgrounds, particularly due to sources of single-electron emission which vary widely across the experiments considered. As detectors improve and the dark rates in the single-electron bin decrease, larger regions of parameter space for DM interacting through plasmon excitation may be uncovered.

Another example of an inelastic detector signal was recently explored by \citet{Ibe:2017yqa} in the context of DM scattering from isolated atoms, known as the Migdal effect. In the standard Migdal effect, orthogonality of initial- and final-state wavefunctions makes the rate proportional to $(m_e/m_N)^2 q^2$, as coherence over the final-state wavefunctions is lost \cite{Ibe:2017yqa, Baxter:2019pnz, Essig:2019xkx}. This could explain why the event rate per unit mass in noble liquid detectors is smaller than in semiconductors, as those amorphous materials lack a pronounced long-range plasmon mode. Furthermore, the parameter space which lies near the thermal relic target is precisely the DM mass range in which the Migdal and direct electron scattering rates are comparable when scattering through a heavy mediator \cite{Baxter:2019pnz,Essig:2019xkx}. 
In fact, in the dark photon model, both processes will be present, giving a markedly different spectral shape to the signal. We emphasize, however, that the isolated atom approximations made in the standard treatment of the Migdal effect fail to take into account long-range interactions in the valence shell that are known to lead to nontrivial collective behavior in solid-state materials; we argue that the dominant signal in the 10--100 MeV mass range in semiconductors is \emph{not} the Migdal effect, but instead plasmon excitation.

\subsection{Scenario 2: Direct Plasmon Excitation Through a Light Mediator}
\label{sec:Scenario2}

Alternatively, we can consider the limit where $m_{A'} \ll q$, such that DM is effectively millicharged. The EELS experiments which characterized plasmons with electron probes can thus be used to determine the event rate for DM-induced plasmons. As mentioned above, DM cannot excite the plasmon directly unless $v \gtrsim 10^{-2}$. Given a velocity distribution with support for $v \gtrsim 10^{-2}$, the plasmon excitation rate per unit detector mass can be derived from the analogous results for EELS. The plasmon excitation probability per incident particle per unit time is \cite{kundmann1988study}
\begin{align}
    P(E) = \frac{\alpha}{\pi^2} \int d^3 & \vecq \, \frac{1}{q^2}{\rm Im}\left \{ \frac{-1}{\epsilon(E, \vecq)} \right \} \nonumber \\
& \times \delta \left(E - \vecq \cdot \vecv + \frac{q^2}{2m_\chi}\right),
\end{align}
where $\epsilon(E, \vecq)$ is the dielectric function of the target. The single delta function enforces energy conservation, but there is \emph{no} corresponding delta function for momentum conservation; it is in this sense that we refer to the plasmon as an inelastic excitation. The plasmon contribution is extracted by considering the region of small momentum transfer and approximating $\epsilon(E, \vecq) \approx \epsilon(E, 0)$, the imaginary part of which gives the plasmon lineshape $S(E)$ (see Appendix~\ref{app:Plasmon} for details). By taking $\alpha \to \kappa^2 \alpha_D$, multiplying by the number of DM particles in the detector volume, and integrating over the velocity distribution and momentum transfer, we obtain the DM-plasmon spectrum per unit detector mass:
\begin{equation}
\label{eq:PlasmonSpectrum}
    \frac{dR}{dE} = \frac{f \rho_\chi}{m_\chi \rho_T} \frac{2 \kappa^2 \alpha_D}{\pi} S(E) \int_{0}^{q_c} \frac{dq}{q} \eta(v_{\rm min}(q,E))~,
\end{equation}
where $\rho_\chi$ is the DM mass density, $\rho_T$ is the target mass density, $\eta(v)$ is the mean inverse DM speed, and 
\be
v_{\rm min}(q,E) = \frac{E}{q} + \frac{q}{2m_\chi},
\ee
is the minimum $\chi$ speed required to deposit energy
$E$. 
Note that we have cut off the $q$ integral at the maximum value of $q_c \sim 2\pi/a \sim 5$ keV compatible with sourcing a long-range plasmon.

The plasmon lineshape $S(E)$ is taken from Ref.~\cite{kundmann1988study} and shown in Fig.~\ref{fig:light_mediator} (left). Following the analysis of  Ref.~\cite{kundmann1988study} for silicon, we normalize $S(E)$ to the Fr\"{o}hlich model of a single damped harmonic oscillator \cite{frohlich1959phenomenological} with core electron dielectric constant $\epsilon_c = 1$ (see Appendix~\ref{app:Plasmon} for further details). To understand the order of magnitude of the rate, we can use the fact that if $\eta(v_{\rm min})$ is approximately independent of $E$, and that in the Fr\"{o}hlich model, $S(E)$ is Lorentzian so 
\be
\label{lorentzian}
\int \frac{dR}{dE} dE \propto \int S(E) \, dE \approx \frac{3}{2} E_p~
\ee
(see Appendix~\ref{app:Plasmon}). This underestimates the true rate slightly because it neglects the long high-energy tail of the germanium plasmon. For a monochromatic velocity distribution at velocity $v$ such that $m_\chi v^2 > E_p$, this gives an approximate total rate
 \be
 \label{eq:ApproxPlasmonRate}
 R \approx \frac{3}{\pi}\frac{f \rho_\chi}{m_\chi \rho_T v}\kappa^2 \alpha_D E_p \log\left(\frac{m_\chi v^2}{E_p}\right)~.
 \ee

In Fig.~\ref{fig:light_mediator} (right) the gray shaded pink region marks parameter space for which the $\chi$-induced direct plasmon excitation yields a $20$ Hz/kg event rate at EDELWEISS, for abundance fractions with $v = 0.1$ ranging from $f = 10^{-5}$ to $10^{-1}$ . The shaded regions of this figure represent astrophysical bounds on millicharged particles, including constraints on $\chi \bar \chi$ emission in red giants \cite{Vogel:2013raa} and supernovae \cite{Chang:2018rso,DeRocco:2019jti}.  The curve labeled ``Freeze-In" represents the parameter space for which {\it the dominant, slower $1-f$} fraction of the $\chi$ population can be produced out of equilibrium through the kinetic mixing interaction \cite{Essig:2011nj,Dvorkin:2019zdi,Hall:2009bx}. The effective DM millicharges $\kappa g_D$ which match the observed rate for $f < 1$ are larger than the millicharges required to generate the observed relic abundance from freeze-in, so for those parameters, some interaction within the dark sector would be required to deplete the DM relic abundance~\cite{Krnjaic:2017tio,Evans:2019vxr}.

Finally, as in Scenario 1, we would expect to see a nonzero rate from the Migdal effect or electron scattering in noble liquid detectors, but one which is smaller than in semiconductors. The excitation rate for a generic system is proportional to $|\langle \Psi_f | e^{i \vecq \cdot \mathbf{x}} | \Psi_i \rangle|^2$, where $\Psi_i$ and $\Psi_f$ are the full many-body electronic wavefunctions. In a solid-state system, these many-body contributions are incorporated in the dielectric function $\epsilon(\vecq, \omega)$ (see Appendix~\ref{app:Plasmon}), and the plasmon represents a many-body state with a very large dipole matrix element $|\langle |\Psi_f | \mathbf{x} |\Psi_i \rangle|^2$, because the wavefunctions have support over many lattice spacings. By contrast, in noble liquids the final-state wavefunction contains an ionized electron, which will not have large overlap with the initial state except in the vicinity of the nucleus. However, the large exposure and the persistent low-energy excesses in xenon and argon experiments listed in Tab.~\ref{tab:eventrates} may still be consistent with a combination of DM-electron scattering and the Migdal effect, as in Scenario 1 above. A more quantitative analysis would also require including the fast DM fraction in the velocity distribution, which we leave for future work.

Regardless of the particular model for the DM velocity distribution with which we choose to compute the rate, we note that plasmon excitation is a striking counterexample to the conventional wisdom that inelastic processes like the Migdal effect dominate at \emph{large} momentum transfers \cite{Baxter:2019pnz,Essig:2019xkx}. While this may be true for isolated atoms, the long-range Coulomb force creates collective excitations which are enhanced at \emph{small} momentum transfer in semiconductors. This example also illustrates the importance of many-body processes which account for electron-electron interactions, as opposed to scattering rates computed using non-interacting single-particle states.

\section{Conclusions}

\label{sec:Conclusions}
In this paper we have argued that multiple  excesses in low-threshold dark matter experiments may be explained by an inelastic excitation, which can be consistently interpreted as a plasmon. We thus predict the following: 
\begin{enumerate}
    \item The ratio of $E_{det}$ to $E_e$ on an event-by-event basis measures the branching fraction of the plasmon to phonons and electron/hole pairs, respectively. The statistical moments of this ratio will be a function of energy, but they should be the same for all events with the same $E_e$ in a given detector material. To our knowledge this branching fraction has not been calculated in the literature; if our interpretation is correct, such a computation would be highly relevant to DM experiments. 
    \item With sufficient resolution (on the order of 1 eV, less than the typical width of the plasmon peak in Si and Ge) and a threshold below the expected plasmon energy, the $E_{\rm det}$ spectrum should show a relative maximum at $E_{\rm det} = E_p$. The $E_e$ spectrum may not show a peak above the $1e^{-}$ bin for charge only detectors. For the CDMS and EDELWEISS detectors run in $E_e$ mode, a significant shift away from the quantized one- and two-electron peaks due to the large excess phonon energy should be observed.
    \item A similar spectrum should be seen in sapphire, where $E_p = 24 \ \eV$, once an energy resolution below 5 eV (approximately the width of the plasmon peak) is achieved.
\end{enumerate}


The striking independence of these excesses with respect to detector composition, location, and environment suggests a common origin. Given the difficulty of explaining plasmon excitation in terms of SM backgrounds, we suggest an interpretation in terms of DM-induced excitations. We have proposed two scenarios, where DM interacts through a heavy or light mediator, producing secondary and primary plasmon excitations, respectively. In the context of these models, we make the following predictions:

\begin{enumerate}
\item Explaining these large signal rates with the dominant halo DM population (as in Scenario 1 of Sec.~\ref{sec:Scenario1}) suggests a 
 single-nucleon contact interaction satisfying $\sigma_p \gtrsim 10^{-35} \ \rm cm^2$ and a DM mass scale $30-200$ MeV. 
 Such large SM couplings and light DM masses imply large DM production rates at terrestrial accelerator searches. In particular, if the underlying interaction is due to an invisibly-decaying dark photon $(m_{A^\prime} > 2m_\chi)$, some combination of Belle-II \cite{Kou:2018nap},  BDX \cite{Battaglieri:2016ggd}, SHiP \cite{Alekhin:2015byh}, NA62 \cite{Mermod:2017ceo},  NA64 \cite{NA64:2019imj}, DUNE \cite{DeRomeri:2019kic}, and LDMX \cite{Akesson:2018vlm} among others will discover or falsify this scenario
 (see also \cite{Battaglieri:2017aum} for a
 broad list of follow-up searches at accelerators).
    \item 
     There should be a reduced annual modulation signal in both Scenarios 1 and 2 (Secs. \ref{sec:Scenario1} and \ref{sec:Scenario2}) compared to the standard expectation from WIMP DM, since DM dominantly deposits energy of order $E_p$ inside the detector, regardless of its initial energy. In particular, the spectrum itself should not show a significant annual modulation, although the overall rate should change due to the modulating DM flux. 
     However, this prediction should be interpreted with great care. For instance, in Scenario 2, the signal arises from a boosted sub-population of the cosmic DM, which could arise from solar reflection \cite{Emken:2017hnp,An:2017ojc} and would not exhibit the expected annual modulation signature at all. Furthermore, even if the source population has a conventional Maxwellian velocity distribution, there are subtleties in interpreting modulation results on sub-annual timescales as the phase of this modulation is sensitive to solar gravitational focusing effects \cite{Lee:2013wza,Lee:2015qva}. Analyzing this effect is beyond the scope of the present work, but may become important in follow-up studies. That said, given the enormous total event rates which have so far been observed, some annual modulation signal should be visible at high statistical significance with enough exposure. 
     \item
     In an anisotropic material where plasmon-phonon interactions or the dielectric function are directional, a daily modulation may be seen. In the direct excitation model, there may also be strong directional signals in low-threshold experiments searching for the dominant $1-f$ cold DM fraction \cite{Hochberg:2017wce,Coskuner:2019odd,Geilhufe:2019ndy,Trickle:2019nya,Griffin:2019mvc}.
     \item The secondary plasmon hypothesis from Sec. \ref{sec:Scenario1} implies a large plasmon-phonon coupling, which may be seen in condensed matter experiments involving neutron energy-loss spectroscopy, for example. 
     \item
     GaAs, a polar material with a direct gap and $E_p \sim 15 \ \eV$, should have even larger plasmon-phonon couplings than Si or Ge and a markedly different branching ratio of the plasmon to phonons compared to Si and Ge, which have indirect gaps. This would result in a larger total signal rate in Scenario 1, with a different relationship between $E_e$ and $E_{det}$ in both Scenarios 1 and 2. Considerable attention has already been devoted to GaAs as a candidate for sub-MeV DM detection \cite{Derenzo:2016fse,Knapen:2017ekk}, and this signals discussed here further motivate investigation of this material. 
\end{enumerate}

A pressing question arising from this analysis is how one might discover or falsify a DM signal which dominantly produces plasmons. In the near future, we believe the most promising line of inquiry would be to operate a detector similar to the EDELWEISS or CDMS HVeV detectors in both calorimetric or charge mode in a low-background environment. To date, no experiment has published results from a detector operating in both modes in an experimental site with known backgrounds; such an experiment could significantly strengthen the case for an inelastic interaction. Furthermore, a signal of this magnitude presents the unique challenge in that it is significantly higher than ambient backgrounds. It is thus important to expose the detector to an elevated background to verify that the observed excess remains unchanged and does not correlate with photon or neutron rates. 

The gold standard of proof beyond these tests, likely at least a few years down the road, is a calorimetric measurement with sufficient resolution and a low enough threshold to detect and resolve the plasmon peak. In addition, it should be demonstrated that rates in different materials should scale according to the strength of the plasmon interaction, and more detailed calculations are needed to reinforce our assertion that the spectrum should closely resemble the plasmon lineshape (which, we stress, can be measured directly with EELS). An important corollary of the conclusions in this paper is that \textbf{charge and light production are secondary processes after the initial energy deposition, and readout of these end-stage signals only provides a relative measurement of the branching ratio of deposited energy into these channels}. Barring a calibration of this branching ratio, results from sub-GeV DM experiments are limited by the uncontrolled systematics related to the assumed charge or light production yield from the primary event.

 Similarly, we emphasize that it is imperative to try to understand the precise rate and spectra of expected signals in the context of liquid noble and scintillation detectors, and to continue to operate these detectors at lower thresholds to gain better insight into the shape of the observed spectra. Upcoming experiments will continue to shed light on the source of dark counts observed by xenon and argon experiments, for example, which will contribute to a better understanding of whether a similarly suggestive rate exists in these experiments. In parallel, more work to measure dynamic structure factors for these materials would elucidate the nature of low-energy inelastic interactions that, at this point, are still not well characterized. An intriguing possibility suggested by Table~\ref{tab:plasmons} is that solid xenon or argon detectors may allow one to test the theory that the event rate is strongly enhanced by the presence of a plasmonic resonance at low energy.

A number of our predictions are nontrivial and represent qualitatively new effects and interpretations of dark matter interactions in condensed matter detectors. We eagerly look forward to the results of upcoming experiments to either support or refute these conclusions. In either case, we expect that the dark matter community will benefit greatly with the increased interactions with the condensed matter community which may be stimulated by this work.

\section*{Acknowledgments}
We gratefully acknowledge Peter Abbamonte for discussions and for pointing out that plasmons are the likely mechanism by which light dark matter couples to charged particles in condensed matter systems, and Lucas Wagner for many enlightening conversations which helped to bridge the gap between the languages of condensed matter physics and high-energy physics. In parallel, we want to acknowledge Alan Robinson and Emile Michaud for pointing out that plasmon interactions should impact low-energy reconstruction of electron recoils. None of the observations in this paper would be possible without the results we cite, but also without private conversations with the collaborations responsible which led to early discussions on the various possible background origins of observed excesses. We thus want to acknowledge (in alphabetical order) Dan Bauer, Karl Berggren, Julien Billard, Alvaro Chavarria, Juan Collar, Rouven Essig, Enectali Figueroa-Feliciano, Jules Gascon, Yonit Hochberg, Ziqing Hong, Tongyan Lin, Sam McDermott, Kaixuan Ni, Paolo Privitera, Matt Pyle, Karthik Ramanathan, Wolfgang Rau, Florian Reindl, Peter Sorenson, Javier Tiffenberg, Belina von Krosigk, and Tien-Tien Yu. We are especially grateful to Nikita Blinov, Torben Ferber, Jeff Filippini, Paddy Fox, Roni Harnik, Dan Hooper, Lauren Hsu, Sam McDermott, Harikrishnan Ramani, Albert Stebbins, and Belina von Krosigk for their feedback on early drafts of this paper. We thank the Gordon and Betty Moore Foundation and the American Physical Society for the support of the ``New Directions in Light Dark Matter'' workshop where the key idea for this work was conceived.  Fermilab is operated by Fermi Research Alliance, LLC, under Contract No. DE-AC02-07CH11359 with the US Department of Energy.  This work was supported in part by the Kavli Institute for Cosmological Physics at the University of Chicago through an endowment from the Kavli Foundation and its founder Fred Kavli.

\appendix

\section{Differentiating Dark Counts from Signal}\label{app:dc}

Part of the difficulty in understanding excesses in single electron experiments is differentiating dark counts (a distinct type of detector background) from a putative signal, given that the rate of dark counts cannot easily be increased as a calibration step in the same way as nuclear or electronic recoil backgrounds. For this reason, it is very hard to interpret the single electron rate as signal. 

We can, however, use the single electron rate to determine whether the higher energy bins are consistent with dark count pileup; if not, we can conclude they arise from a distinct source. Given a single electron dark rate $\Gamma_d$ and an integration window $T$, we find a mean number of dark events $\lambda_d = \Gamma_d T$. For measurements which integrate a fixed window of time, such as CCDs, we should therefore get a Poisson distribution with mean $\lambda_d$ of events, as in~\cite{PhysRevLett.123.181802}.

CCDs actually sample all pixels, even empty ones, so those pixels without dark counts show up in the event histogram. We thus can determine a Poisson mean from the 0 and 1 electron bins, and predict the dark rate in the second bin; an excess above this rate is from a different source. For example, for a dark rate of 400 Hz/kg as seen in the SENSEI data 1-electron bin, a mass of about $4 \times 10^{-10}$~kg (the mass of a single pixel), and an exposure of an hour, we find that the mean $\lambda_d$ is roughly $5\times 10^{-4}$. This means that we expect the second bin to have a rate of roughly 0.1 Hz/kg, which is more than an order of magnitude lower than the measured rate. We can therefore conclude that the excess in the second bin is likely from a different source than the single electron bin, or that these two bins are not entirely dominated by dark counts.

For experiments which take a time-stream of data, we find that the probability of a dark event being irreducible pileup on top of another dark event is $p_{pu} = \Gamma_d\delta t$, where $\delta t$ is the minimum separation between events that can be distinguished. For example, in CDMS HVeV with a minimum time discrimination of $10 \mu s$ \cite{Agnese_2018} and a dark rate of 1 Hz in a 1g detector, we thus expect that $p=10^{-5}$. For a measured dark rate of $10^3$~Hz/kg, we expect the pileup rate to be $10^{-2}$~Hz/kg. This is $10^3$ times lower than the measured rate in any of the 2-6 electron bins; we can therefore conclude that if the first bin is truly all dark counts, none of the other bins can possibly be dark counts unless the experiment is observing a correlated leakage process. Because models of dark counts only produce a single charge at a time, we conclude that the event rates in the charge bins above the single electron bin are not due to dark counts.

This analysis allows us to state the following: while the single electron bin in both SENSEI and CDMS HVeV may be dominated by leakage, we can exclude simple leakage as being the source of events in the $>$1 electron bins. There must be, at the least, another unknown source of events. In the main text, we follow this logic to exclude SM particles as the source, motivating a DM interpretation.

\section{Using Electron Yield to Determine Recoil Type}\label{app:recoil}

Conventional DM searches, probing physics at energy scales above $\sim$1~keV, classify events in a binary fashion as electron recoil (ER) or nuclear recoil (NR). This clean separation is due to the very different microphysics involved in each, and the large number of additional interactions which occur during the relaxation process as a high-energy particle deposits its momentum and energy in the detector.

In the case of a semiconductor, an electron recoil produces a pair of high-energy charges, which interact primarily with other electrons, efficiently generating electron-hole pairs at a mean rate of $n_{eh}=E_e/\epsilon_{eh}$, where $n_{eh}$ is the number of generated electron-hole pairs, $E_e$ is the initial energy in the electron system, and $\epsilon_{eh}$ is the mean energy per electron-hole pair produced. A remarkably generic property of semiconductors is that $\epsilon_{eh}$ is constant across energy scales from eV to MeV (see e.g. Ref.~\cite{Canali72}); $\epsilon_{eh}\sim3.6$~eV for Si and 3.0~eV for Ge, for example. This provides a convenient energy scale for charge detectors measuring $n_{eh}$, under the assumption that all events measured are interaction with electrons. The electron-equivalent energy scale $E_e$, measure in eVee, is just $n_{eh}\epsilon_{eh}$.

Nuclear recoils, on the other hand, involve a more complex picture. At very high energies (well above 15 eV in Si and Ge \cite{chen1969orientation}), a nuclear recoil produces a crystal defect: a nucleus is physically removed from its lattice site and bounces through the lattice, displacing some nuclei, ionizing others, and producing phonons. The relative scattering rate for ionization versus other loss processes is thus much more complex than electron scattering; in particular, screening effects, which depend on the momentum of the nucleus, become important. This means the relative energy given to the electron system, $E_e$, versus the phonon system, $E_r$, is momentum-dependent. At energies above 1~keV, this energy partition is well-modeled by a charge screening model, the so-called Lindhard model \cite{Agnese_2017,Barker_2013}, but at low energies experiments have begun to see sharp departures from this smooth energy dependence (see e.g. Ref.~\cite{Scholz_2016}).

\begin{figure}[t]
    \centering
    \includegraphics[width=0.45\textwidth]{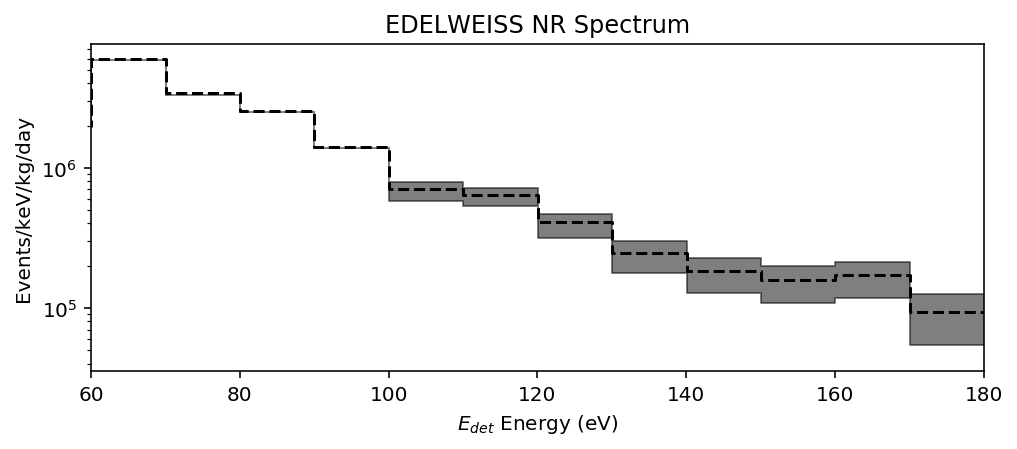}
    \includegraphics[width=0.45\textwidth]{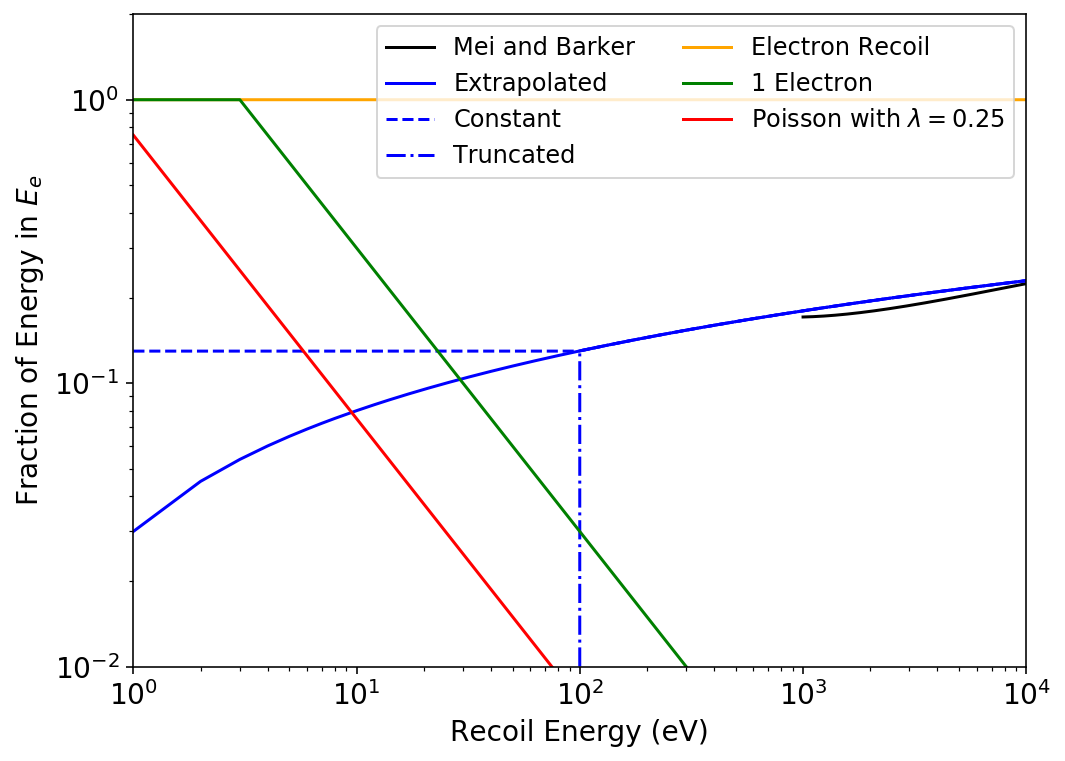}
    \caption{\textbf{Top:} EDELWEISS excess in $E_{det}$ \cite{EdelweissWIMP} used for the various $E_e$ spectrum models in Figure~\ref{fig:Rates}. The grey band shows the uncertainty in the measurement; bins near threshold have very small uncertainties due to the large number of events observed below 100~eV. \textbf{Bottom:} Yield curves used to generate simulated $E_e$ spectra in Figure~\ref{fig:Rates}. The electron recoil (ER) yield curve is flat because energy is initially imparted to the $E_e$ system, and thus $E_e=E_{det}$. The NR curves show a decreased yield at lower energy. We have extended the NR yield model from Ref~\cite{Agnese_2017} down to the lowest calibrated energy; we then consider the full range of options from an extrapolation, to constant yield, to yield which discontinuously goes to 0. We also show the model from Ref~\cite{Barker_2013} for comparison. Finally, we show two inelastic yield curves which assume a fixed charge mean produced, one with a Poisson distribution (red), the other with identically one charge produced per event (green).}
    \label{fig:yield}
\end{figure}

The key point is that, if two of the three quantities $(E_e, E_r, E_{det})$ can be measured (usually one proportional to $E_e$, the other either $E_r$ or $E_{det}=E_r+E_e$), then the recoil type of a given event can generally be determined using the ratio of these quantities by comparing to calibrated yield curves. This is generally applicable at energies above $\sim$1~keV, and has been used successfully by all recent DM experiments searching for DM with masses above a few GeV. Below these energies, however, better resolution is required in both measurements to successfully discriminate by recoil type, and statistical fluctuations begin to wash out discrimination ability on an event by event level. For this reason, many of the very low threshold experiments have resorted back to only measuring one quantity. In our analysis, we are primarily comparing experiments which only measure charge, and thus report spectra in $E_e$, with experiments which measure heat, and are thus sensitive to $E_{det}=E_e+E_r$. All electron recoil DM searches, by construction, are only measuring $E_e$.

The lack of a clear distinction between ER and NR at sub-keV energies becomes important when we try to compare two experiments in this regime in the presence of some unknown signal. In this paper, we consider the excesses observed by EDELWEISS \cite{EdelweissWIMP,edelweissHV} because, having measured both $E_e$ and $E_{det}$ with similar detectors, we are able to statistically determine the mean yield as a function of energy for a population of events, and thereby shed light on the origin of those events. Consider the $E_e$ spectrum shown in the different panels of Figure~\ref{fig:Rates} compared with the $E_{det}$ spectrum plotted in Figure~\ref{fig:yield}. Ideally, both measurements would be made with infinite precision, and without a finite threshold. The remarkable development of single charge detectors over the past few years allows the $E_e$ measurement to have effectively no threshold, but the $E_{det}$ measurement still has a threshold $E_{thresh} = 60 \ \eV$.

We can still, however, try to determine whether events are electron recoil, nuclear recoil, or an entirely different class of events by applying the yield model and ensuring that, when we convert the spectrum from $E_{det}$ to $E_e$, the result is not in direct conflict with the measured spectrum. Put another way, we must have
\begin{equation}
 R_{total} = \int_0^{\infty} \frac{dR}{dE_e}dE_e > \int_{E_{thresh}}^{\infty}\frac{dR}{dE_{det}}dE_{det}.
\end{equation}

We thus convert the $E_{det}$ spectrum to a modeled $E_e$ spectrum as
\begin{equation}
    \frac{dR_{model}}{dE_e} = y(E_{det})^{-1}\frac{dR}{dE_{det}}
\end{equation}
where for this paper, this is a convolution done by Monte Carlo. We then check that 
\begin{equation}\label{eq:ruledout}
    \int_{E_{low}}^{E_{high}}\frac{dR_{model}}{dE_e} < \int_{E_{low}}^{E_{high}}\frac{dR_{meas}}{dE_e}
\end{equation}
for any choice of $E_{low}$ and $E_{high}$. If we could sample all of the $E_{det}$ spectrum, this would be an equality, but because we know we're only sampling the $E_{det}$ spectrum above $E_{thresh}$, the model should always either match or undershoot the measured $E_e$ spectrum, which is complete.

The yield models we consider in this paper, shown in Figure~\ref{fig:yield}, constitute a constant yield (ER), falling yield (NR), and fixed $E_e$ independent of $E_{det}$. Because the yield is a ratio of $E_e$ to $E_{det}$, these last models appear to rise in yield space at low energy. One of the primary conclusions of this paper is that the only yield models which satisfy Eq.~(\ref{eq:ruledout}) for the measured $E_e$ spectrum in Figure~\ref{fig:Rates} are the rising models; in other words, the spectra are inconsistent with either a NR or ER interaction, and suggest a novel inelastic interaction. This also illustrates that interpreting either the $E_e$ spectrum or $E_{det}$ spectrum independently, without knowledge of the type of recoil, leads to erroneous exclusion curves. For example, the NR limit implied by converting the $E_e$ spectrum to an effective nuclear recoil energy scale would be overly aggressive.

\section{Plasmon Review}
\label{app:Plasmon}

Since plasmons in solid-state systems are likely unfamiliar to many high-energy physics, in this appendix we review some basic properties of plasmons and their measurement using EELS. To facilitate comparison with the literature, we will use as much as possible the notation of Ref.~\cite{kundmann1988study}, in contrast to the typical high-energy physics notation of the main text (i.e. $\omega$ instead of $E$, and $e$ instead of $\alpha$).

\subsection{Plasmon measurements with EELS}

Plasmons are the quantized longitudinal oscillations of valence electrons in a condensed matter system, carrying energy on the order of the classical plasma frequency. In EELS experiments, plasmons are excited by the electric field of an electron traversing the material, which has a longitudinal component (i.e. there is a component of the field along the electron's direction of motion). The response of a material to electromagnetic fields of momentum $\vecq$ and frequency $\omega$ can be characterized by a complex dielectric function $\epsilon(\omega, \vecq)$. For an electron with charge $e$, mass $m_e$, and velocity $v$ traversing a material with dielectric function $\epsilon$, the differential probability per unit time of depositing energy $\omega$ is \cite{kundmann1988study}
\begin{align}
\label{eq:dPdtdomega}
\frac{dP}{dt d\omega} = \frac{e^2}{4\pi^3} \int d^3 & \vecq \, \frac{1}{q^2}{\rm Im}\left \{ \frac{-1}{\epsilon(\omega, \vecq)} \right \} \nonumber \\
& \times \delta \left(\omega - \vecq \cdot \vecv + \frac{q^2}{2m_e}\right),
\end{align}
where we have converted to the Heaviside-Lorentz units conventional in high-energy physics (in contrast to the formulas from \cite{kundmann1988study} which use Gaussian units common in condensed matter physics and contain additional factors of $4\pi$).\footnote{Note that in the small-$q$ limit, the $q^2/(2m_e)$ term is negligible, so this term is typically neglected in the condensed matter literature when considering electron probes.} Note that we can rewrite the energy conservation condition enforced by the delta function as 
\be
q = \frac{E}{v\cos \theta} + \frac{q^2}{2m_e v \cos \theta}
\ee
where $\theta$ is the angle between $\vecq$ and $\vecv$. We are interested in the forward-scattering region where $\cos \theta > 0$ where momentum transfer is the smallest. In that part of phase space, both terms on the right-hand side are positive-definite, so we obtain the inequality $q \geq E/v$ by dropping the second term. Assuming the plasmon has typical energy $\omega = E_p$, we find the important relation
\be
\label{eq:PlasmonConditionqApp}
q \geq \frac{E_p}{v},
\ee
which is saturated in the forward scattering limit where $\vecq$ is parallel to $\vecv$ and when the finite-mass term $\frac{q^2}{2m_e}$ is negligible (which typically holds for the kinematics relevant to EELS). This condition is simply an expression of energy conservation, which must be satisfied to excite the plasmon at the peak energy. Define $q_p = E_p/v$ as the typical scale of momentum transfer for plasmon excitations. As mentioned in the main text, the plasmon has a cutoff frequency $q_c \sim 2\pi/a$. Plasmon resonances have been observed with $q$ about a factor of 2 above this cutoff \cite{raether2006excitation}, but for parametric estimates, it will suffice to require that $q_p < q_c$. To satisfy Eq.~(\ref{eq:PlasmonConditionqApp}), we must have
\be
\label{eq:PlasmonConditionvApp}
v \geq E_p/q_p = 6.5 \times 10^{-3} \left(\frac{E_p}{16 \ \eV}\right)
\ee
Note that this condition is \emph{independent} of the mass of the incident particle; it could be an electron, proton, or millicharged DM. In the main text we conservatively tighten this bound on $v$ to $10^{-2}$. This constraint on the velocity explains why EELS experiments to probe the plasmon are performed with semi-relativistic electrons.

Assuming the plasmon kinematic condition (\ref{eq:PlasmonConditionvApp}) is satisfied, the presence of the $1/q^2$ in the integrand of Eq.~(\ref{eq:dPdtdomega}) implies that the smallest allowed momentum transfers, $q \sim q_p$, will dominate. This is just the typical behavior of the long-range Coulomb force. In that case, we can approximate $\epsilon(\omega,\vecq) \approx \epsilon(\omega, 0)$ and pull it out of the $\vecq$ integral, giving
\be
\label{eq:PlasmonProbFixedV}
\frac{dP}{dt d\omega} = \frac{e^2}{2\pi^2 v}{\rm Im}\left \{ \frac{-1}{\epsilon(\omega)} \right \} \log \left(\frac{2E_0}{\omega}\right).
\ee
where $E_0$ is the incident electron energy. The logarithmic dependence on $E_0$ is a manifestation of the universal Coulomb logarithm, which here is cut off by the energy transfer $\omega$. This formula can be modified in a straightforward way for relativistic probes. Note that since $v = \sqrt{2E_0/m_e}$, the plasmon excitation probability scales as $\log(E_0)/\sqrt{E_0}$, a relation which holds for any nonrelativistic charged particle (in particular, for millicharged DM as well as electrons). 

Note that by dividing by $v$ and integrating over $\omega$, we can convert this expression into a probability per unit length for the probe to undergo some nonzero energy loss. Setting this to unity gives an inelastic mean free path, which for electrons ($Q = e$) with $v = 0.1$ is about 58 nm for 50 keV electrons in silicon \cite{raether2006excitation}. For electrons, then, multiple scattering is an important consideration, and in thick samples this will give rise to energy deposits in integer multiples of $E_p$. On the other hand, for DM with a small millicharge, the mean free path scales with $\kappa^2 g_D$ and is orders of magnitude larger than the detector size for the millicharge values we consider. Thus, multiple scattering of DM will typically not occur in small semiconductor detectors. 

\subsection{Role of the dielectric function}

In the non-interacting electron approximation, the complex dielectric function of a material is given by the Lindhard formula~\cite{dressel}:
\begin{widetext}
\begin{equation}
\epsilon(\omega,\vecq) = 1 - \lim_{\eta \to 0} \frac{e^2}{V}\frac{1}{\vecq^2} \sum_{\veck} \sum_{l,l'}  \frac{f_{\rm FD}(E_{\veck + \vecq, l'}) - f_{\rm FD}(E_{\veck,l})}{E_{\veck + \vecq, l'} - E_{\veck,l} - \omega - i \eta} \times  |\langle |\veck + \vecq, l' | e^{i \vecq \cdot \vecx} | \veck, l\rangle|^2.
\label{eq:Lindhard}
\end{equation}
\end{widetext}
Here, $V$ is the volume of the material, $l$ and $l'$ are band indices, $E_{\veck,l}$ is the energy of the $l^\text{th}$ band at lattice momentum $\veck$, and $f_{\rm FD}$ are Fermi-Dirac factors. The imaginary part of $\epsilon$ reflects ``on-shell'' transitions when two states have an energy difference $\omega$. The dielectric function is closely related to the 1-loop vacuum polarization in quantum field theory, which has similar properties (i.e. its imaginary part reflects on-shell final states). Note that ${\rm Im}\{-1/\epsilon\} = {\rm Im} \, \epsilon/|\epsilon|^2$ selects out these transitions and weights them by the corresponding squared matrix element, which is why this expression appears in the formula Eq.~(\ref{eq:dPdtdomega}). 

It is important to emphasize that Eq.~(\ref{eq:Lindhard}), which forms the basis for much of the recent literature on DM interactions in condensed matter systems \cite{Essig:2015cda,Hochberg:2015fth,Hochberg:2017wce}, assumes no electron-electron interactions.  In this approximation, \emph{the plasmon does not appear.} In the language of high-energy physics, the plasmon is analogous to a pole in a matrix element which does not correspond to the fields in the Lagrangian.\footnote{Strictly speaking, one can understand the plasmon \emph{energy} from the Lindhard function by treating the electrons as a free Fermi gas and taking the limit $\vecq \to 0$, giving $\epsilon(\omega, 0) = 1 - \omega_p^2/\omega^2$; however, this approximation does not give an imaginary part to the dielectric function and hence does not explain the finite width $\Gamma$.} This is by no means unusual, and indeed is behavior characteristic of strongly-coupled field theories. Moreover, much like the Sudakov factor in QCD, the plasmon is a result of a resummation of an infinite series of diagrams (known in the condensed matter literature as the random phase approximation \cite{fetter2012quantum}) and does not appear at any finite order in perturbation theory. Thus is is best to treat the plasmon phenomenologically as a nonperturbative effect which nonetheless contributes to the imaginary part of the dielectric function. 

Finally, we note that the factor of $e^2$ in the dielectric function represents the response of a material to (ordinary, electromagnetic, possibly virtual) photons. The DM-induced direct plasmon excitation rate is proportional to $\kappa^2 \alpha_D$, which roughly speaking represents the probability of millicharged DM emitting a virtual dark photon which converts to a virtual photon. The dielectric function then parameterizes the response of the material to this photon, which is proportional to $e^2$ in perturbation theory.\footnote{We note in passing that if DM-SM interactions were mediated by a light scalar rather than a light vector, a completely different object (essentially a scalar response function) would control the material response. This function has been calculated in perturbation theory for the case of isolated atoms \cite{Catena:2019gfa}, but it would be very interesting to understand the nonperturbative plasmon contribution in solid-state systems in light of our work, which cannot be measured with any SM probe because no long-range scalar forces exist in the SM.}

\subsection{Plasmon lineshape}
The Fr\"{o}hlich damped-harmonic-oscillator model \cite{frohlich1959phenomenological} posits the following form for $\epsilon(\omega)$:
\be
\epsilon(\omega) = \epsilon_c + \frac{E_p^2}{(E_g^2 - \omega^2) - i \Gamma \omega},
\ee
where $\epsilon_c$ is the contribution to the dielectric constant from core electrons (assumed independent of $\omega$), $E_p$ is the plasma energy of the valence electrons, $E_g$ is an average band gap, and $\Gamma$ is the plasmon damping parameter. As in the analogous Breit-Wigner formulae in high-energy physics, $\Gamma$ represents the sum of the partial widths of all the plasmon decay modes, including to phonons and electron/hole pairs. In high-energy physics, if the decay of a resonance of energy $E$ can be described perturbatively, the narrow-width approximation $\Gamma/E \ll 1$ applies. For plasmons, typical values are $\Gamma = 3 \ \eV$ and $E = 16 \ \eV$ \cite{raether2006excitation, kundmann1988study}, so $\Gamma/E \approx 0.2$ is not particularly small. For comparison, $\Gamma/E$ for the top quark is 0.008, and $\Gamma/E$ for the lowest-lying $\Delta$ resonance is 0.09, so in this sense the plasmon is even more nonperturbative than QCD resonances. Thus it is not unreasonable that the plasmon-phonon coupling, which controls both plasmon production and plasmon decay, is nonperturbatively large, consistent with Scenario 1 in the main text.

Substituting into Eq.~(\ref{eq:PlasmonProbFixedV}), we find
\be
\label{eq:Frohlich}
\frac{dP}{dt d\omega} = \frac{e^2}{2\pi^2 v} \frac{\Gamma E_p^2 \omega \ln(2E_0/\omega)}{\epsilon_c^2(E_g^2 + E_p^2/\epsilon_c^2 - \omega^2)^2 + \Gamma^2 \omega^2}.
\ee

This model is an excellent fit to the observed plasmon in silicon, but the germanium plasmon has a longer high-energy tail due to contributions from the core $3d$ electrons. However, the region around the peak is well-modeled by a single Lorentzian, so following \cite{kundmann1988study} we take $\epsilon_c =1$ and use Eq.~(\ref{eq:Frohlich}) to normalize the germanium plasmon, taking the best-fit values in the two-parameter model of \cite{kundmann1988study} with an effective plasmon energy $E_p'^2 \equiv E_g^2 + E_p^2/\epsilon_c^2$. Integrating over $\omega$ (or equivalently $E$) for millicharged DM gives Eq.~(\ref{eq:ApproxPlasmonRate}). On the other hand, to obtain the spectrum for a general velocity distribution $f(v)$, we weight Eq.~(\ref{eq:dPdtdomega}) by $f(v)$ and integrate over $v$. Solving the delta function by performing the velocity integral, as is standard in DM-electron scattering treatments, and performing the $q$ integral up to $q_c$ gives Eq.~(\ref{eq:PlasmonSpectrum}) in the main text. In that equation, the (dimensionless) plasmon lineshape is
\be
S(\omega) \equiv {\rm Im}\left \{ \frac{-1}{\epsilon(\omega)} \right\},
\ee
which is
\be
\label{eq:SFrohlich}
S_F(\omega) \equiv \frac{\Gamma E_p^2 \omega}{\epsilon_c^2(E_g^2 + E_p^2/\epsilon_c^2 - \omega^2)^2 + \Gamma^2 \omega^2}
\ee
in the Fr\"{o}hlich model. Integrating Eq.~(\ref{eq:SFrohlich}) gives
\be
\int S_F(\omega) \, d\omega \approx \frac{3}{2}\epsilon_c E_p
\ee
for $E_g = 0$.

\bibliography{dm}

\end{document}